\documentclass[
reprint,
superscriptaddress,
 amsmath,amssymb,
 aps%,
]{revtex4-2}

\usepackage[pdftex]{graphicx}
\usepackage[pdftex]{xcolor}
\usepackage[breaklinks,pdftex]{hyperref}

\begin{document}

\title{
Spin Seebeck Effect as a Probe for Majorana Fermions in Kitaev Spin Liquids
}

\author{Yasuyuki~Kato}
\email{katoyasu@u-fukui.ac.jp}
\affiliation{Department of Applied Physics, University of Fukui, Fukui 910-8507, Japan}
\affiliation{Department of Applied Physics, The University of Tokyo, Tokyo 113-8656, Japan}

\author{Joji~Nasu}
\affiliation{Department of Physics, Tohoku University, Sendai, Miyagi 980-8578, Japan}

\author{Masahiro~Sato}
\affiliation{Department of Physics, Chiba University, Chiba 263-8522, Japan}

\author{Tsuyoshi~Okubo}
\affiliation{Institute for Physics of Intelligence, The University of Tokyo, Tokyo 113-0033, Japan}

\author{Takahiro~Misawa}
\affiliation{Beijing Academy of Quantum Information Sciences, Haidian District, Beijing 100193, China}
\affiliation{The Institute for Solid State Physics, The University of Tokyo, Kashiwa, Chiba 277-8581, Japan}

\author{Yukitoshi~Motome}
\affiliation{Department of Applied Physics, The University of Tokyo, Tokyo 113-8656, Japan}

\date{\today}

\begin{abstract}
Quantum entanglement in strongly correlated electron systems often leads to exotic elementary excitations.
Quantum spin liquids provide a paradigmatic example, where the elementary excitations are described by fractional quasiparticles such as spinons. 
However, 
such fractional quasiparticles behave differently from electrons, 
making their experimental identification challenging.
Here, we theoretically investigate the spin Seebeck effect, 
which is a thermoelectric response via a spin current, as an efficient probe of the fractional quasiparticles in quantum spin liquids, 
focusing on the Kitaev honeycomb model. 
By comprehensive studies using the real-time dynamics, 
the perturbation theory, and the linear spin-wave theory based on the tunnel spin-current theory, 
we find that the spin current is induced by thermal gradient in the Kitaev spin liquid, via the low-energy fractional Majorana excitations. 
This underscores the ability of Majorana fermions to carry spin current, despite lacking spin angular momentum. 
Furthermore, we find that the induced spin current changes its sign depending on the sign of the Kitaev interaction,  
indicating that the Majorana fermions 
contribute to the spin current with (up-)down-spin-like nature when the exchange coupling is (anti)ferromagnetic. 
Thus, in contrast to the negative spin current already found in a one-dimensional quantum spin liquid, 
our finding reveals that the spin Seebeck effect can exhibit either positive or negative signals, 
contingent upon the nature of fractional excitations in the quantum spin liquids.
We also clarify contrasting field-angle dependence between the Kitaev spin liquid in the low-field limit and the high-field ferromagnetic state, 
which is useful for the experimental identification.
Our finding suggests that the spin Seebeck effect could be used not only to detect fractional quasiparticles emerging in quantum spin liquids but also to generate and control them.
\end{abstract}

\maketitle

%%%%%%%%%%%%%%%%%%%%%%%%%%%%%%%%%%%%%%%%%%%%%%%%%%%%%%%%%%%%%%%
\section{Introduction}
%%%%%%%%%%%%%%%%%%%%%%%%%%%%%%%%%%%%%%%%%%%%%%%%%%%%%%%%%%%%%%%
Quantum entanglement is the key to the emergence of topological phases which have been of central interest in strongly correlated electron systems.
The topological phases can be regarded as vacuums of fractionalized elementary excitations 
which behave differently from the original electrons.
A representative example is the fractional quantum Hall states in two-dimensional electron systems under a magnetic field~\cite{Tsui1982}, 
where electrons and magnetic fluxes form composite elementary excitations with fractional charges obeying neither the Bose-Einstein nor Fermi-Dirac statistics, 
called non-Abelian anyons~\cite{Laughlin1983,Arovas1984,Halperin1984}.
Numerous studies have been conducted to identify and manipulate the non-Abelian anyons~\cite{Feldman2021}, 
since they are expected to be useful for future quantum technologies, such as fault-tolerant quantum computing~\cite{Kitaev2003,Nayak2008}.

Quantum spin liquids, initiated by the resonating valence bond (RVB) state proposed by P. W. Anderson~\cite{Anderson1973}, 
are another interesting platform for fractional excitations.
In the RVB state, charge degree of freedom of electrons is frozen, 
and instead the spin degree of freedom is fractionalized into the so-called spinon and vison~\cite{Kivelson1989,Read1989,Wen1991,Senthil2000,Moessner2001}. 
The RVB state was originally proposed as a candidate for the ground state of the antiferromagnetic (AFM) Heisenberg model on a two-dimensional triangular lattice, 
although it was later found not to be the ground state~\cite{Huse1988,Bernu1992}. 
Nevertheless, it has subsequently led to dramatic developments in the study of high-temperature superconductivity~\cite{Anderson1987,Baskaran1987,Affleck1988,Wen1996} 
and frustrated magnetism~\cite{Balents2010,Lacroix2011,Savary2017}. 
Through these developments, although some important concepts such as topological order were established~\cite{Wen2004}, 
a complete understanding of the quantum spin liquids has not yet been achieved.

The Kitaev spin liquid (KSL) is a rare quantum spin liquid that has been rigorously shown to be the ground state of a realistic quantum spin model on a two-dimensional honeycomb lattice~\cite{Kitaev2006}.
The exact solution allows one to scrutinize the fundamental properties of fractional excitations. 
In the KSL, spins are fractionalized into 
itinerant Majorana fermions and localized $Z_2$ fluxes at zero magnetic field, and these two form 
a composite obeying anyonic statistics under a magnetic field.
Stimulated by the proposal for realization of the model~\cite{Jackeli2009}, tremendous efforts have been devoted to
exploring candidate materials~\cite{Takagi2019,Motome2020,Trebst2022}. 
In addition, close collaborations between experiment and theory have unveiled exotic properties of the KSL~\cite{Winter2017,Knolle2019b,Takagi2019,Janssen2019,Motome2020b}. 
A striking finding is the half-integer quantized thermal Hall effect as a direct evidence of the fractional Majorana excitations~\cite{Kasahara2018,Yokoi2021}, 
while the experimental results are still controversial~\cite{Yamashita2020,Bruin2022,Lefrancois2022,Czajka2023}. 
It was pointed out that the sample quality matters~\cite{Kasahara2022}. 
Significant contributions from phonons were also discussed~\cite{Ye2018,Vinkler-Aviv2018}.
Recently, there are several theoretical proposals for generating, detecting, and controlling the fractional excitations in the KSL, 
e.g., by using the scanning tunneling microscopy~\cite{Carrega2020,Feldmeier2020,Pereira2020,Konig2020,Udagawa2021}, 
interferometers~\cite{Aasen2020,Klocke2021,Klocke2022},
impurity effects~\cite{Knolle2019,Kao2021}, 
local lattice distortions~\cite{Jang2021}, 
and magnetic fields~\cite{Harada2023}, 
whose experimental demonstrations are awaited.

%%%%%%%%%%%%%%%%%%%%%%%%%%%%%%%%%%%%%%%%%%%%%%%%%%%%%%%%%%%%%%%
\begin{figure}[!h]
%\fcolorbox{red}{white}{
\includegraphics[trim = 0 0 0 0, width=\columnwidth]{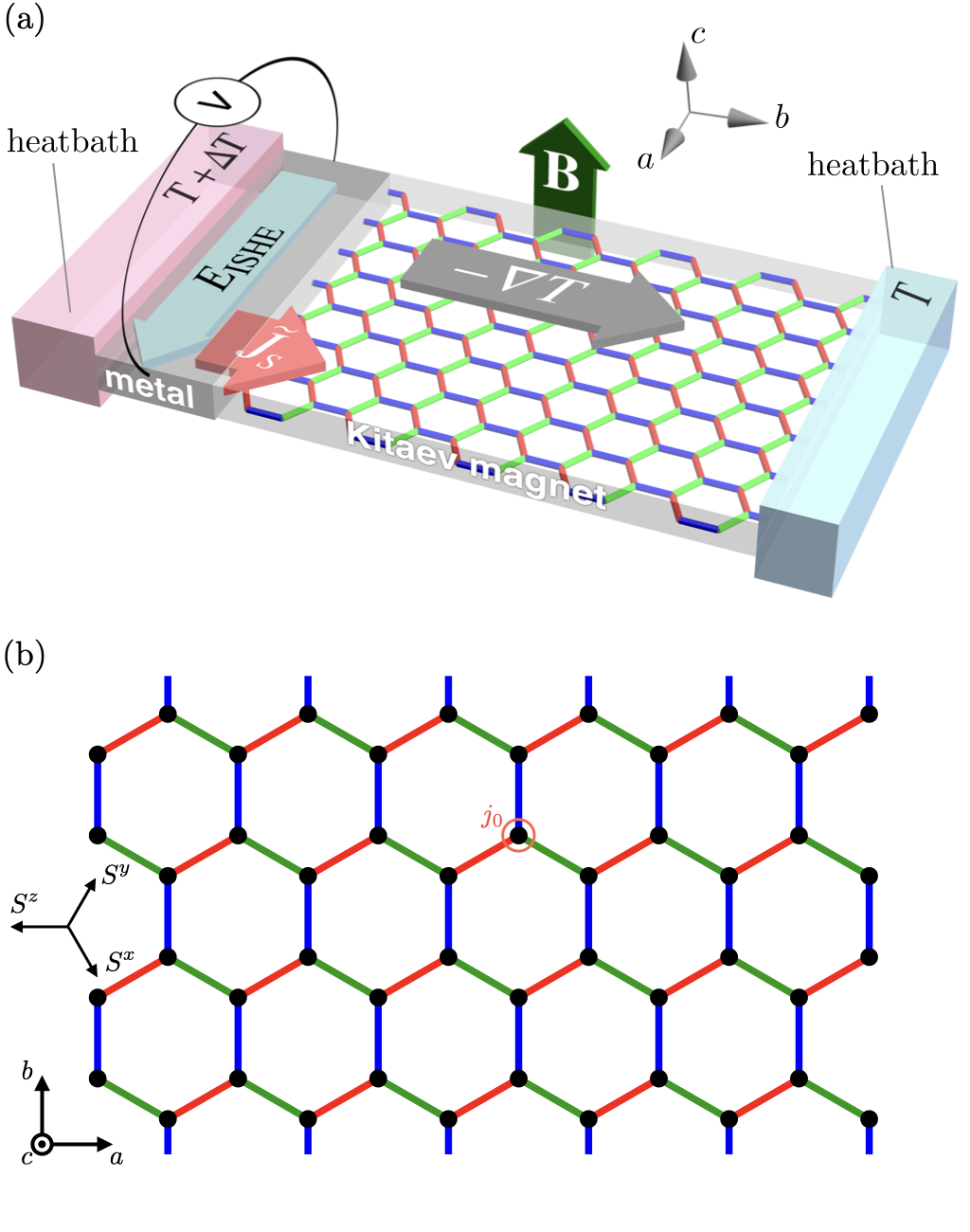}
%}
\caption{\label{fig:fig01}
(a) Schematic of an experimental setup for measuring the spin Seebeck effect in Kitaev magnets.
A Kitaev magnet is attached to heatbaths from two sides to generate a temperature gradient ($-\nabla T$).
A metal with strong spin-orbit coupling is sandwiched at one end. 
Due to the inverse spin Hall effect, the spin current is converted into an electric field $E_{\rm ISHE}$ (light blue arrow) perpendicular to the thermal gradient in the metal, which is observable.
The tunnel spin current $\tilde{J}_S$ (red arrow) and 
the magnetic field ${\bf B}$ parallel to $c$ (green arrow) are also depicted.
We define the sign of $\tilde{J}_S$ to be positive when 
the tunnel spin current polarized antiparallel to ${\bf B}$ direction flows from the hot side to the cold side.
Actual direction of the tunnel spin current depends on the sign of the Kitaev interaction; see Sec.~\ref{sec:results}.
(b) Kitaev model on a honeycomb lattice used in simulations.
The black dots represent the spin-1/2 degrees of freedom, and the red, green, and blue lines represent the $x$, $y$, and $z$ bonds, respectively~[see Eq.~\eqref{eq:H}].
The lattice with $X=4$ and $L=12$ is shown, where $X$ is the number of spins in the vertical direction and $L$ is that in the horizontal direction.
We impose an open boundary condition in the horizontal direction and a periodic boundary condition in the vertical direction.
The $a$, $b$, and $c$ axes and $S^x$, $S^y$, and $S^z$ axes are shown in the inset.
$j_0$ denotes the site where the local dynamical spin susceptibility is calculated. 
}
\end{figure}
%%%%%%%%%%%%%%%%%%%%%%%%%%%%%%%%%%%%%%%%%%%%%%%%%%%%%%%%%%%%%%%

The spin Seebeck effect (SSE) refers to a phenomenon in which a spin current is induced by a thermal gradient.
This effect is widely used as a sensitive probe for 
elementary excitations in magnets~\cite{Seki2015,Wu2016,Geprags2016,Shiomi2017,Li2019,Li2020,Wei2020,Kikkawa2023}. 
In fact, in ordinary ferromagnets and antiferromagnets, 
the SSE is induced by magnon excitations from the conventional magnetic ordered state. 
Recently, this probe has been extended to more nontrivial magnets, such as those with frustrated 
interactions and low dimensionality, 
whose elementary excitations may have both positive and negative spin angular momenta, 
or even become spinless. 
The representative example is a negative SSE observed 
in a quantum spin liquid in a quasi-one-dimensional antiferromagnet~\cite{Hirobe2017}. 
This peculiar behavior, in contrast to conventional ferromagnets, 
was ascribed to fractional spinon excitations forming the Tomonaga--Luttinger liquid. 
Besides, other systems with strong quantum fluctuations, such as 
a spin-nematic liquid~\cite{Hirobe2019}, 
a spin-Peierls system~\cite{Chen2021},
and a magnon Bose--Einstein condensation system~\cite{Xing2022}, 
have also been reported to exhibit the peculiar SSE.
The SSE was also theoretically predicted for the KSL~\cite{Takikawa2022}. 
However, this is caused by chiral edge modes in the KSL, 
and no study of bulk SSE for quantum spin liquids 
in more than one dimension has been reported thus far to the best of our knowledge.
Recently, unconventional spin transport properties to thermal conductivity 
mediated by Majorana excitations in the KSL have been reported experimentally~\cite{Hirobe2017b}
and theoretically~\cite{Minakawa2020,Aftergood2020,Misawa2023}. 
Nonetheless, it remains unclear how the SSE, which involve both spin and heat transport, 
is manifested in the KSL and whether it is useful to identify, generate, and control the fractional excitations.

In this paper, we theoretically analyze the spin current induced by a thermal gradient via the SSE in the KSL. 
The possible experimental setup is shown in Fig.~\ref{fig:fig01}(a); 
we consider an interface between a Kitaev magnet and a paramagnetic metal, and calculate the tunnel spin current through the interface
by the local dynamical spin susceptibility in an applied magnetic field, 
based on the tunnel spin-current theory~\cite{Jauho1994,Adachi2011}.
Accurately calculating the dynamics of quantum many-body problems, particularly in more than one spatial dimension, poses a significant challenge. 
In addressing this issue, we opt for a state-of-the-art numerical technique, 
real-time dynamics simulations based on the time-dependent variational principle (TDVP)~\cite{Haegeman2011}
with matrix product states (MPSs)~\cite{Fannes1992,Schollwock2011}.
We also complementarily conduct the analysis based on the perturbation theory with respect to the magnetic field for the KSL, as well as the linear spin-wave theory for the field-induced ferromagnetic (FFM) state.
As a result, we find that the spin current is induced by a thermal gradient in the KSL 
mediated by the low-energy fractional Majorana quasiparticles that have no spin angular momentum. 
This is qualitatively different from the fact that 
fractional spinons in forming a one-dimensional Tomonaga--Luttinger liquid induce a negative SSE
since the spinon has spin-$1/2$.
It is remarkable that the quasiparticles without angular momentum can contribute to the tunnel spin-current transport.
Furthermore, we find that
the sign of the tunnel spin current depends on the sign of the Kitaev interaction; 
namely, the ferromagnetic (FM) and AFM KSLs lead to a positive and negative SSE, respectively.
This is in stark contrast to the behavior in conventional ferromagnetic states where 
the magnon excitations lead to a positive SSE. 
In addition, we find that the KSLs show distinctive field angle dependence of the spin current that is different from those of the FFM states.
These peculiar behaviors, which are ascribed to the spin current mediated by fractional Majorana excitations, 
are revealed for the first time to our knowledge by using the state-of-the-art numerical technique. 
In turn, our findings suggest the possibility of generating and controlling the Majorana excitations by spin injection.

The structure of this paper is as follows.
In Sec.~\ref{sec:model_and_method}, we describe the model and method used in this study.
After introducing the Kitaev model under an external magnetic field, 
we describe the tunnel spin-current theory
and numerical methods of real-time dynamics simulations 
to compute the SSE.
We also mention the choice of model parameters and the details of calculation conditions.
In Sec.~\ref{sec:results},
we present the results of the real-time dynamics simulations, 
the perturbation analysis, and the linear spin-wave theory. 
In Sec.~\ref{sec:discussion}, we discuss the contrasting behaviors of the tunnel spin current depending on the sign of the Kitaev interaction, 
and the strength and direction of the magnetic field, 
which are important for experimental confirmation, and the experimental implication and observability of the SSE in the KSL.
Finally, Sec.~\ref{sec:summary} is devoted to summary.

%%%%%%%%%%%%%%%%%%%%%%%%%%%%%%%%%%%%%%%%%%%%%%%%%%%%%%%%%%%%%%%
\section{Model and method}\label{sec:model_and_method}
%%%%%%%%%%%%%%%%%%%%%%%%%%%%%%%%%%%%%%%%%%%%%%%%%%%%%%%%%%%%%%%
We consider the Kitaev model~\cite{Kitaev2006} under an external magnetic field on a honeycomb lattice. 
The Hamiltonian reads
\begin{align}
\mathcal{H} = K
\sum_{\gamma}\sum_{\langle j,j' \rangle_\gamma} S^\gamma_j S^\gamma_{j'}
 - \sum_j {\bf B} \cdot {\bf S}_j,
\label{eq:H}
\end{align}
where the first term represents the bond-dependent Ising-type Kitaev interaction with the coupling constant $K$  
and the second term denotes the Zeeman coupling with the magnetic field ${\bf B}$.
The sums of $\gamma$ and $\langle j,j' \rangle_\gamma$ run over $\{x,y,z\}$ and all the $\gamma$ bonds, respectively 
[see Fig.~\ref{fig:fig01}(b)];
${\bf S}_j=(S^x_j, S^y_j, S^z_j)$ represents the spin-$1/2$ operators at $j$ site.
The $g$ factor and the Bohr magneton $\mu_{\rm B}$ are omitted in the Zeeman coupling term.
We take $K =-1$ and $+1$ for the FM and AFM couplings, respectively.

We investigate the SSE in the Kitaev model in Eq.~\eqref{eq:H}
based on the tunnel spin-current theory~\cite{Jauho1994,Adachi2011,Chen2021,Masuda2023}.
The experimental setup is shown in Fig.~\ref{fig:fig01}(a); the SSE is measured by the inverse spin Hall effect~\cite{Saitoh2006,Valenzuela2006,Kimura2007}
 in a metal with strong spin-orbit coupling, such as Pt, attached to a Kitaev magnet. 
In the tunnel spin-current theory, 
spin injection is assumed to occur locally through the interface between the metal and the magnet, 
and calculated under the following conditions:
(i) The interaction between the metal and the magnet is small enough to be dealt with as a perturbation and conserves angular momenta,
(ii) the polarization of the tunnel spin current is parallel to the direction of the magnetic field, 
(iii) the temperature difference between the metal and the magnet is small, 
and (iv) the metal is nonmagnetic with featureless density of states in the low-energy region.
It is worth noting that 
although the total magnetization is not conserved in the Kitaev model, 
these assumptions allow us to evaluate the tunnel spin current transferred to the metal that is detected in experiments. 

On these assumptions, the tunnel spin current induced by a thermal gradient is evaluated by the perturbation theory with respect to the interface interaction using the nonequilibrium Green function method~\cite{Haug2008,Stefanucci2013,Zagoskin2016}. 
The tunnel spin current per one interface site is given by 
\begin{align}
\tilde{J}_S 
=& 
\int_{-\infty}^{\infty}
d\omega \, k
\left( \frac{\omega}{T} \right) \,
 {\rm Im} \chi^{-+}_{\rm loc} (\omega),
\label{eq:Js_org} 
\end{align} 
up to a constant coefficient proportional to the square of the interface interaction and the temperature difference between the metal and magnet. 
In this paper, the sign of the tunnel spin current is defined such that $\tilde{J}_S>0$ when the tunnel spin current polarized antiparallel to the magnetic field direction flows in the direction of the red arrow in Fig.~\ref{fig:fig01}(a).
The local dynamical spin susceptibility of the magnet, $\chi^{-+}_{\rm loc}(\omega)$, is
defined as 
\begin{align}
\chi^{-+}_{\rm loc} (\omega) := -  i  \int_0^\infty
dt
\langle 
\big[
S^-_{j_0}(t),S^+_{j_0}(0)
\big]
\rangle
e^{  i  \omega t - \delta t},
\label{eq:chi-+}
\end{align}
with the damping factor $\delta$; 
$S^\pm_{j_0}$ is defined according to the direction of the magnetic field, namely,
\begin{align}
S^{\pm}_{j_0} = S_{j_0}^{\tilde{x}}\pm i S_{j_0}^{\tilde{y}},
\end{align}
where $\tilde{x}$ and $\tilde{y}$ represent the directions perpendicular to the magnetic field along $\tilde{z}$ (the $\tilde{x}\tilde{y}\tilde{z}$ coordinate forms a right-handed system)~\footnote{
We define $j_0$ on only one of the two sublattices as shown in Fig.~\ref{fig:fig01}(b) and neglect the sublattice dependence 
because trial simulation results do not show recognizable difference in $\chi_{\rm loc}^{-+}(\omega)$ when $j_0$ is defined on the other sublattice.
}. 
In Eq.~\eqref{eq:Js_org}, the integral kernel $k( x=\omega/T )$ is given by 
\begin{align}
k(x) = 
\frac{ 
x^2
 }{\sinh^{2}\big( x/2  \big)},
\end{align}
with the temperature $T$ and the angular frequency $\omega$ (the Boltzmann constant and the Dirac constant are set to be unity). 
This factor $k( \omega/T )$ stems from the dynamical susceptibility of the metal and the temperature difference at the interface~\cite{Adachi2011,Masuda2023}.
Since $k( \omega/T )$ is an even function of $\omega$, Eq.~\eqref{eq:Js_org} is written as
\begin{align}  
\tilde{J}_S 
=& 
\int_{-\infty}^{\infty}
d\omega \,
k 
\left(
\frac{\omega}{T}
\right)\,
[X_+(\omega) +X_-(\omega)] \nonumber \\
=& 
\int_{-\infty}^{\infty}
d\omega\, 
k
\left(
\frac{\omega}{T}
\right)\,
 X_{+}(\omega) ,
\label{eq:Js}
\end{align}
where
$X_{\pm}(\omega)$ are 
the symmetric and antisymmetric components of  ${\rm Im}\chi^{-+}_{\rm loc}(\omega)$, 
defined as
\begin{align}
X_{\pm}(\omega) = \frac{1}{2} {\rm Im}\big[\chi^{-+}_{\rm loc} (\omega) \pm \chi^{-+}_{\rm loc} (-\omega)  \big].
\label{eq:X_pm}
\end{align}

%%%%%%%%%%%%%%%%%%%%%%%%%%%%%%%%%%%%%%%%%%%%%%
\begin{figure}[htp]

\includegraphics[trim = 0 0 0 0, width=\columnwidth, clip]{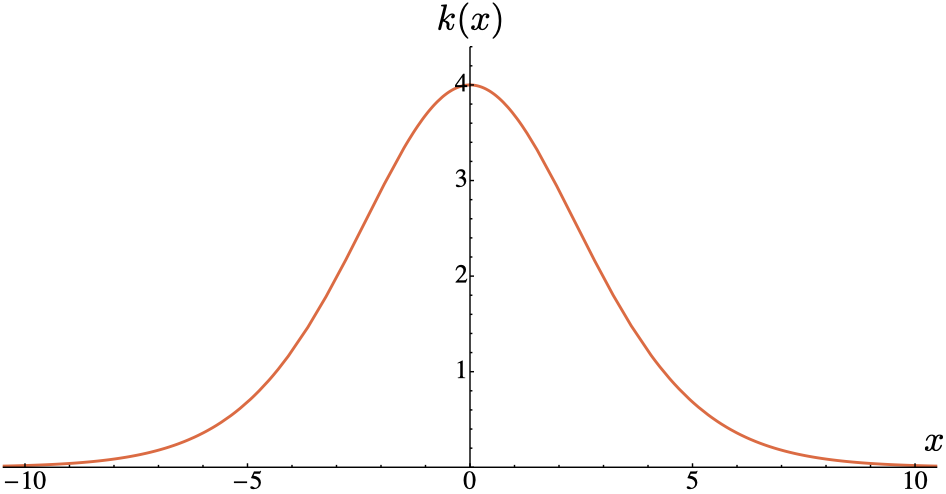}

\caption{\label{fig:fig02}
$x(=\omega/T)$ dependence of the integral kernel $k(x)$. 
}
\end{figure}

Equation~\eqref{eq:Js} shows that only the symmetric component of ${\rm Im}\chi^{-+}_{\rm loc}(\omega)$ contributes to the tunnel spin current. 
The kernel $k( \omega/T )$ has a peak at $\omega=0$ and decays quickly for large $|\omega|/T$; hence, it works as a low-pass frequency filter at each $T$.
Figure~\ref{fig:fig02} shows the $\omega/T$ dependence 
of $k( \omega/T )$. 
The relevant frequency range is given by $|\omega / T| \lesssim 5$, 
and it becomes wider for higher temperature; 
for instance,
at $T=0.01$, $X_+(\omega)$ for $|\omega| \lesssim 0.05$ is relevant to the tunnel spin current, 
and at $T=0.04$, $X_+(\omega)$ for $|\omega| \lesssim 0.2$ is relevant. 
We note that the tunnel spin current is always positive in the high-temperature limit, since the kernel becomes constant as 
$\lim_{T\to \infty}k( \omega/T )=4$ and 
the sum rule $\int_0^\infty X_+(\omega) d\omega = 2 \pi M$ holds ($M$ is the average of spin moment in the field direction): 
$\lim_{T \to \infty}\tilde{J}_S =8\pi M$.

%%%%%%%%%%%%%%%%%%%%%%%%%%%%%%%%%%%%%%%%%%%%%%
\begin{figure}[htp]
\includegraphics[trim = 0 0 0 0, width=\columnwidth, clip]{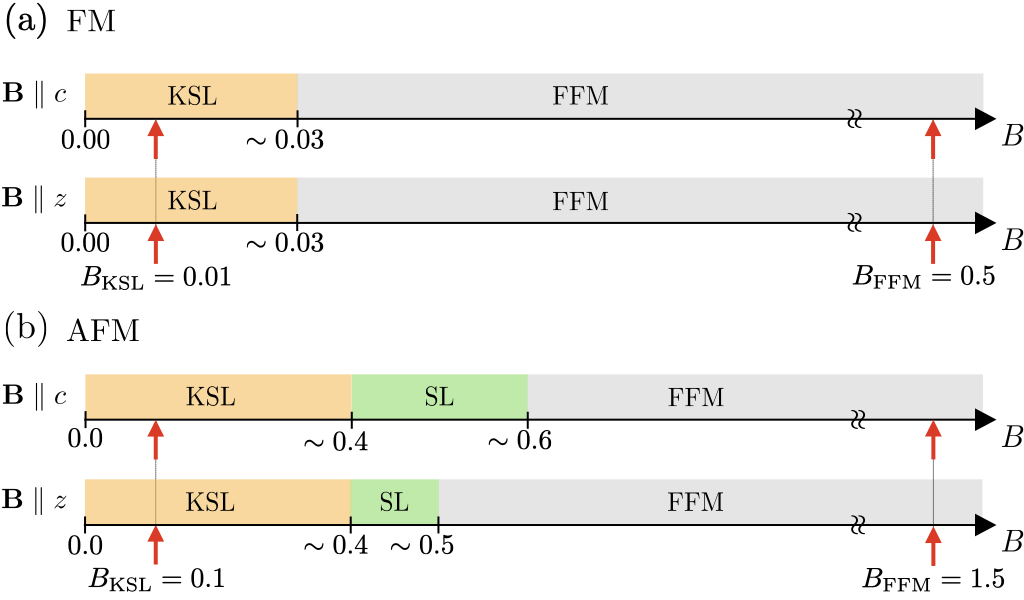}
\caption{\label{fig:fig03}
Schematic ground-state phase diagrams for the Kitaev model under the magnetic field $B$
 for the (a) FM and (b) AFM cases.
The magnetic field directions are ${\bf B}\parallel c$ in the upper panel and ${\bf B}\parallel z$ in the lower panel in each figure. 
The magnetic fields used in the present calculations, $B_{\rm KSL}$ and $B_{\rm FFM}$ are indicated by the red arrows.
The values of the critical fields are taken from Refs.~\cite{Jiang2011,Zhu2018,Nasu2018,Hickey2019}.
}
\end{figure}

In this study, we compute the tunnel spin current in an applied magnetic field ${\bf B}$ along the  $\tilde{z}(=z$, $a$, $b$, or $c )$ axis [see Fig.~\ref{fig:fig01}(b)], and mainly 
discuss the cases of ${\bf B}\parallel z$ and ${\bf B}\parallel c$;
the former case gives the Zeeman coupling as $-BS_j^z$, 
while the latter gives $-B(S_j^x+S_j^y+S_j^z)/\sqrt{3}$ in the second term in Eq.~\eqref{eq:H}.
In the calculations of $X_+(\omega)$,
we choose $S_{j_0}^{\pm}$ as 
$S_{j_0}^{\pm} = S_{j_0}^{x}\pm  i S_{j_0}^{y}$ for ${\bf B}\parallel z$, and
$S_{j_0}^{\pm} = S_{j_0}^{a}\pm  i S_{j_0}^{b}$ for ${\bf B}\parallel c$ 
(cyclic permutations for ${\bf B}\parallel a$ and ${\bf B}\parallel b$).
We investigate the low-field KSL as well as the high-field FFM state for comparison.
The phase transition from the KSL to the FFM state was studied 
by using the exact diagonalization, 
the density matrix renormalization group (DMRG)~\cite{White1992,Schollwock2005}, 
and Majorana mean-field approximation~\cite{Jiang2011,Zhu2018,Nasu2018,Hickey2019}.
The phase diagram is different between the FM and AFM cases 
not only quantitatively but also qualitatively, as shown in Fig.~\ref{fig:fig03}. 
In the FM case, the KSL turns into the FFM state at 
$B \approx 0.03$ for both ${\bf B}\parallel c$~\cite{Jiang2011,Zhu2018,Hickey2019} and ${\bf B}\parallel z$~\cite{Nasu2018}; 
there is a single phase transition. 
Meanwhile, in the AFM case, the values of the critical field become one order of magnitude larger, 
and moreover, the system exhibits another phase transition in addition to that to the FFM state.
For ${\bf B}\parallel c$, 
the KSL is stable up to $B\approx 0.4$, 
and a different spin liquid (SL) takes place before entering into the FFM state at $B\approx 0.6$~\cite{Zhu2018,Hickey2019}.
For ${\bf B}\parallel z$,
the KSL is stable up to $B\approx 0.4$, 
and the intermediate SL is predicted to be stable up to $B\approx 0.5$~\cite{Nasu2018}.
With reference to these studies,
we choose the values of $B$ for the KSL (FMM) state, $B_{\rm KSL}$ ($B_{\rm FFM}$), well below (above) the critical fields: 
$B_{\rm KSL} = 0.01$ and $B_{\rm FFM} = 0.5
$ for the FM case, and 
$B_{\rm KSL} = 0.1$ and $B_{\rm FFM} = 1.5
$ for the AFM case, as indicated by the red arrows in Fig.~\ref{fig:fig03}. 
%%%%%%%%%%%%%%%%%%%%%%%%%%%%%%%%%

In the following calculations, 
we compute $\chi^{-+}_{\rm loc}(\omega)$ in Eq.~\eqref{eq:chi-+} at zero temperature for simplicity. 
This leaves the temperature dependence of the tunnel spin current only in the kernel $k(\omega/T)$. 
This assumption should be justified at sufficiently low temperature
where the $Z_2$ flux excitations are scarce, typically
below the crossover temperature $T\approx 0.016$ 
at zero field~\cite{Nasu2015}. 
To compute $\chi^{-+}_{\rm loc}(\omega)$ at $T=0$, 
we perform a real-time dynamics simulations based on TDVP~\cite{Yang2020}.
The initial states of the simulations are obtained by multiplying $S^\pm_{j_0}$ to the MPS ground states 
computed by DMRG with the bond dimension $400$.
We consider a cylinder-shaped lattice structure with the periodic (open) boundary condition in the vertical (horizontal) direction, 
as shown in Fig.~\ref{fig:fig01}(b), 
of the system size $X\times L$ with $(X,L)=(4,12)$, $(4,16)$, $(6,6)$, $(6,8)$, and $(6,10)$, 
where $X$ and $L$ are the number of sites in the vertical and horizontal directions, respectively, 
and employ the MPS representation wrapping around the cylinder in a  snake form.
We note that the ground state is not degenerate 
for all cases at the chosen magnetic fields. 
In Eq.~\eqref{eq:chi-+}, 
we choose $j_0$ at the site close to the center of each cylinder, 
as shown in Fig.~\ref{fig:fig01}(b), and take $\delta=0.01$.
We set one time step of real-time dynamics as $\Delta t =0.2$ and perform $N_t = 10^4$ time evolutions: 
Simulations are performed up to a maximum time $t_{\rm max}=2000$.
When performing the Fourier transform~[Eq.~\eqref{eq:chi-+}], 
we interpolate the data by a cubic spline.
The calculations are performed using a software library iTensor~\cite{itensor}.
In addition,
to understand the simulation results,
we analyze the dynamical spin susceptibility based on the perturbation theory 
with respect to the magnetic field and the linear spin-wave theory, 
as described in Secs.~\ref{sec:perturb} and \ref{sec:spinwave}, respectively.

\section{Results}\label{sec:results}
%%%%%%%%%%%%%%%%%%%%%%%%%%%%%%%%%%%%%%%%%%%%%%%%%%%%%%%%%%%%%%%
%%%%%%%%%%%%%%%%%%%%%%%%%%%%%%%%%%%%%%%%%%%%%%%%%%%%%%%%%%%%%%%

\subsection{Real-time dynamics simulations}\label{sec:realtime}

\begin{figure}[!htp]
\centering
\includegraphics[trim = 0 0 0 0, clip,width=\columnwidth]{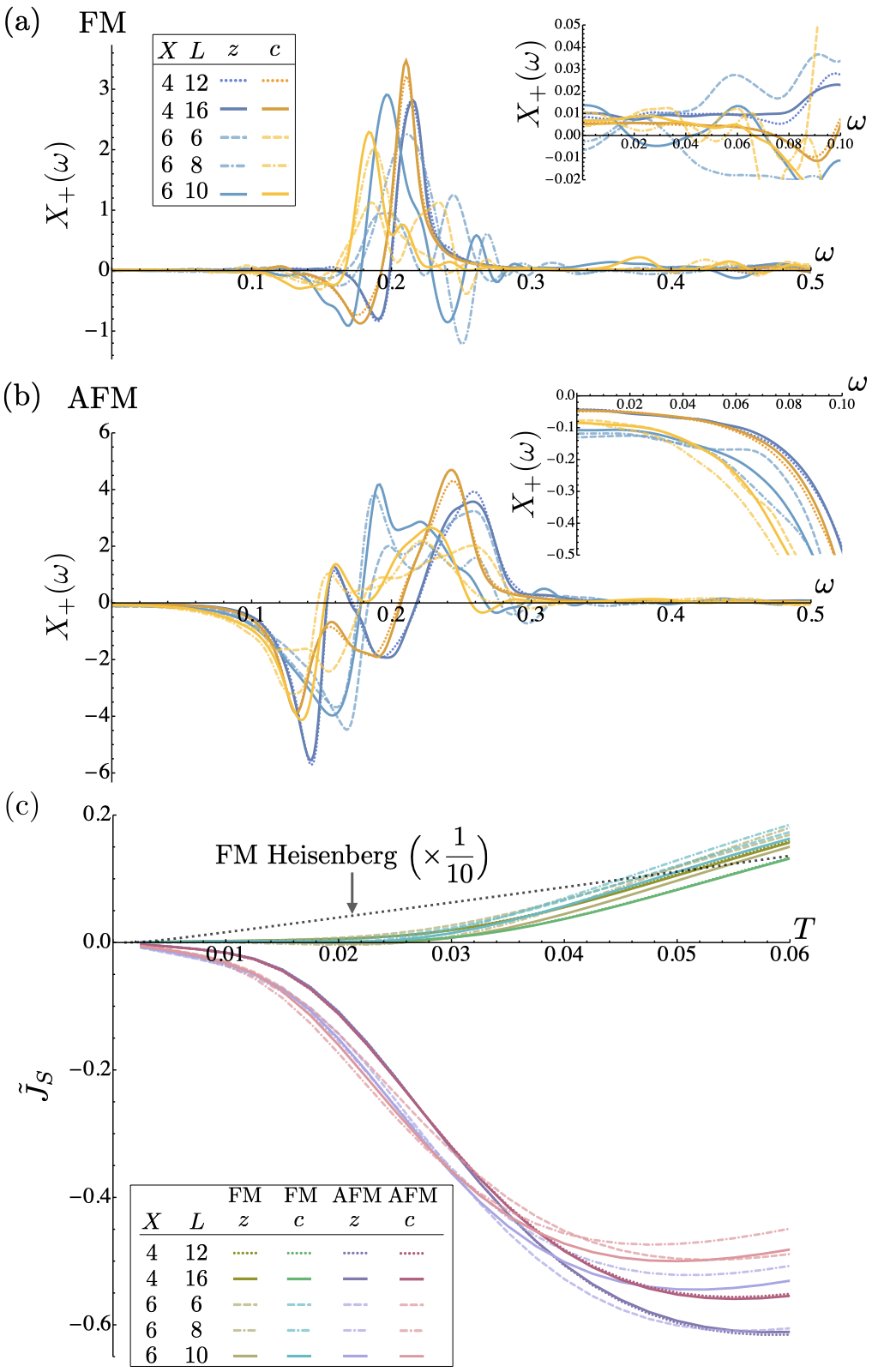}

\caption{\label{fig:fig04}
Results of real-time dynamics simulations in the KSL at $B=B_{\rm KSL}$. 
(a) and (b) show the $\omega$ dependences of $X_+(\omega)$ [Eq.~\eqref{eq:X_pm}] for the FM and AFM Kitaev models, respectively.
The insets are the enlarged plots in the low-frequency region. 
The different line colors and types represent the different field directions and the system sizes, respectively.
(c) shows the temperature dependences of the tunnel spin current $\tilde{J}_S$ [Eq.~\eqref{eq:Js}]. 
The black dotted line in (c) is a reference data for the $S=1/2$ FM Heisenberg model at $B=0.01$, 
multiplied by a factor of $1/10$ for better visibility (see Appendix~\ref{app:A}).
}
\end{figure}
\begin{figure}[!htp]
\centering

\includegraphics[trim = 0 0 0 0, clip,width=\columnwidth]{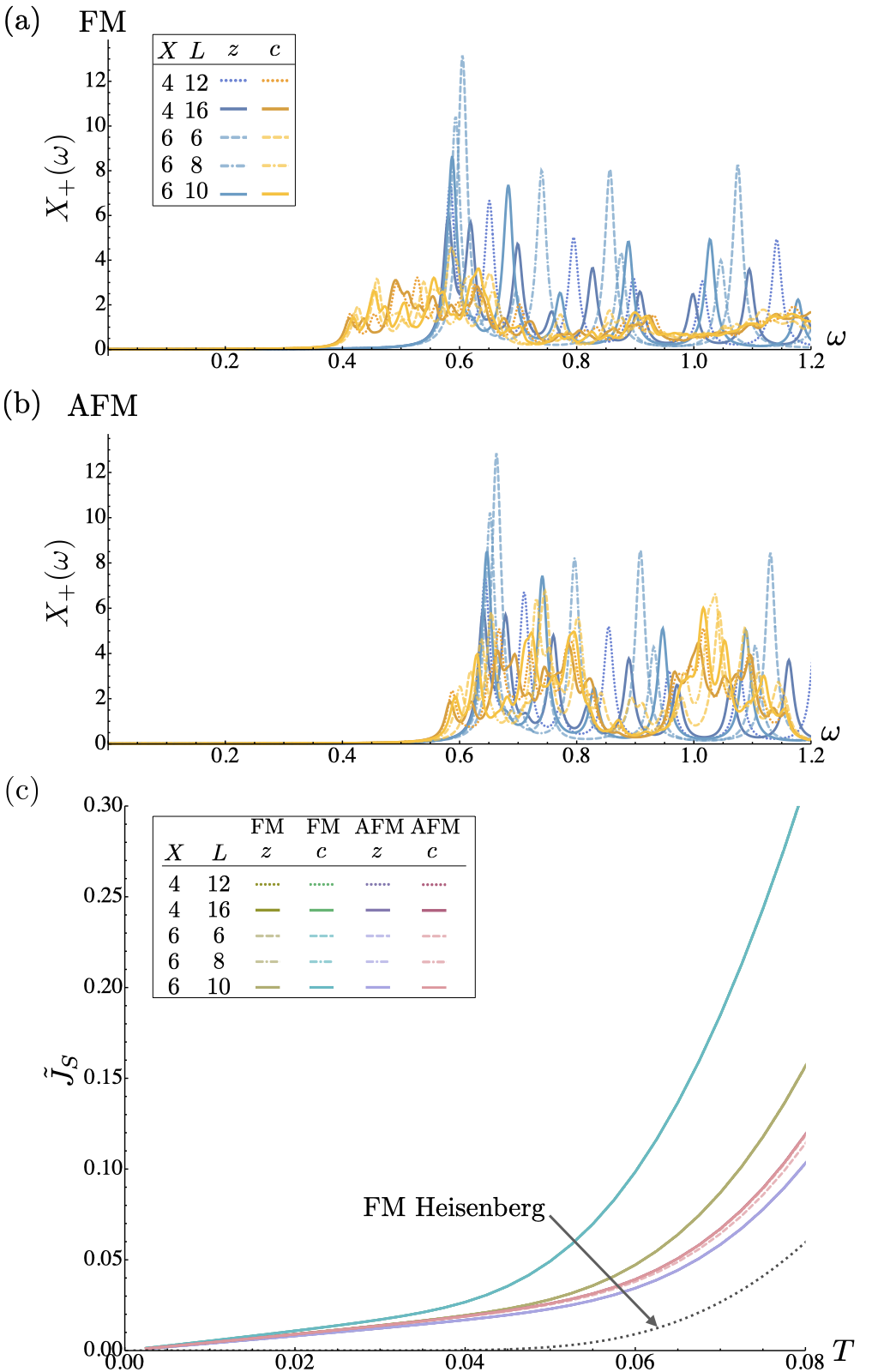}

\caption{\label{fig:fig05}
Results in the FFM state at  $B=B_{\rm FFM}$.
The notations are common to those in Fig.~\ref{fig:fig04}.
The black dotted line in (c) is a reference data for the $S=1/2$ 
FM Heisenberg model at $B=0.5$ (see Appendix~\ref{app:A}). 
}
\end{figure}

We present the results of the real-time dynamics in the KSL at a low magnetic field $B=B_{\rm KSL}$ (see Fig.~\ref{fig:fig03}) 
obtained by TDVP in Fig.~\ref{fig:fig04}. 
Figures~\ref{fig:fig04}(a) and \ref{fig:fig04}(b) show the symmetrized dynamical spin susceptibility $X_+(\omega)$ [Eq.~\eqref{eq:X_pm}]
for the FM and AFM Kitaev models, respectively, and Fig.~\ref{fig:fig04}(c) summarizes the tunnel spin current $\tilde{J}_S$ [Eq.~\eqref{eq:Js}]. 
First of all, we show that, in both FM and AFM KSL states, the tunnel spin current is induced at finite temperature. 
In addition, we find a clear difference in the temperature dependence between the FM and AFM cases: $\tilde{J}_S$ is suppressed at low $T$, experiencing growth with a positive value as $T$ increases for the FM case, but it takes on a largely negative value in the AFM case. 
This trend is consistently observed across different system sizes and geometries. 
The origin can be traced back to the $\omega$ dependence of $X_+(\omega)$. 
In the FM case, $X_+(\omega)$ is small in the low-$\omega$ regime; the data are overall positive, but still scattered due to the finite-size effects, as shown in Fig.~\ref{fig:fig04}(a). 
In the larger-$\omega$ region, $X_+(\omega)$ exhibits a large positive intensity after a small negative dip, leading to the positive growth of $\tilde{J}_S$ with increasing $T$ in Fig.~\ref{fig:fig04}(c). 
In contrast, in the AFM case, $X_+(\omega)$ is developed with a negative value from the low- to intermediate-$\omega$ regime irrespective of the system sizes as shown in Fig.~\ref{fig:fig04}(b), which contributes to the largely negative $\tilde{J}_S$ in Fig.~\ref{fig:fig04}(c). 
Notably, while the finite-size effects are noticeable in $X_+(\omega)$ for the clusters accessible by the state-of-the-art numerical technique, those pertaining to $\tilde{J}_S$ are relatively small, showing the definite trend in the temperature dependence. 
We therefore conclude that, in the KSL state under the magnetic field, the tunnel spin current is thermally induced with opposite signs depending on the sign of the Kitaev interaction $K$.

This is in stark contrast to the behaviors in the FFM state at a high field $B=B_{\rm FFM}$. 
The results are shown in Fig.~\ref{fig:fig05}. 
In this case, $X_+(\omega)$ is overall positive, including the low-$\omega$ region, for both FM and AFM cases, as shown in Figs.~\ref{fig:fig05}(a) and \ref{fig:fig05}(b). 
This leads to the tunnel spin current with positive sign for both cases, as shown in Fig.~\ref{fig:fig05}(c).
These behaviors are confirmed by the linear spin-wave theory in Sec.~\ref{sec:spinwave}.

The contrasting behaviors between the KSL and FFM states can be ascribed to the fact that the carriers of the spin current are different between the two states. 
In the high-field FFM state, the elementary excitations are conventional magnons, 
which give a positive spin current irrespective of the details of the magnetic coupling.
In contrast, in the low-field KSL state, the spin degree of freedom is fractionalized into itinerant Majorana fermions and localized $Z_2$ fluxes.
They are no longer independent of each other under magnetic fields, but in a low-field and low-temperature regime, 
the spin current in the KSL is expected to be carried predominantly by the itinerant Majorana fermions. 
The remarkable sign difference of $\tilde{J}_S$ between the FM and AFM Kitaev models suggests that the nature of
Majorana fermions is different between the FM and AFM KSLs. 
This is a qualitatively different result from the one-dimensional antiferromagnet,
where the spinon-mediated spin current was found to be negative~\cite{Hirobe2019}.
We will discuss this sign switching by using the Majorana representation and the perturbation theory with respect to the magnetic field in Sec.~\ref{sec:perturb}.

Let us make two remarks. 
The first one is on the magnitude of the tunnel spin current, $|\tilde{J}_S|$. 
In the KSL, $|\tilde{J}_S|$ increases much faster with temperature in the case of AFM than FM. 
Since the field strength $B$ is different and the low-field phase diagrams look different between the two cases, 
it is difficult to conclude that this faster increase of $|\tilde{J}_S|$ is a general tendency. 
Nevertheless,
$|\tilde{J}_S|$ are sufficiently large to be observed in experiments in both FM and AFM cases. 
This is explicitly shown by the comparison with the prototypical $S=1/2$ FM Heisenberg model, whose tunnel spin current is plotted by the black dotted line in Fig.~\ref{fig:fig04}(c) 
(see Appendix~\ref{app:A} for the details).
Note that the spinon-mediated 
spin current was observed in the previous experiment~\cite{Hirobe2017},
even though it is supposed to be smaller than the magnon-mediated one by a factor of $10^{-4}$--$10^{-3}$.
We will further discuss in Sec.~\ref{sec:exp_obs} the experimental observability by quantitative comparison with a typical three-dimensional (3D) ferromagnet in a realistic experimental setup.
Meanwhile, in the FFM state, $|\tilde{J}_S|$ is almost the same at low temperature regime in FM and AFM. 
This is due to almost the same magnitude of Zeeman gap, as shown in Figs.~\ref{fig:fig05}(a) and \ref{fig:fig05}(b) for the FM and AFM cases, respectively, 
which is further confirmed by the linear spin-wave theory in Sec.~\ref{sec:spinwave}.
Figure~\ref{fig:fig05}(c) also shows $\tilde{J}_S$ for the FM Heisenberg model at $B=0.5$ for comparison; 
the magnitude is much smaller than the result for $B=0.01$ shown in Fig.~\ref{fig:fig04}(c) 
because of the larger Zeeman gap, 
but it is comparable to those for the FFM states in the FM and AFM Kitaev models.

The second remark is on the dependence on the field direction.
In Fig.~\ref{fig:fig04}(c), the tunnel spin currents in the low-field KSL show almost the same temperature dependences for ${\bf B}\parallel z$ and ${\bf B}\parallel c$. 
We confirm that the results for ${\bf B}\parallel a$ and ${\bf B}\parallel b$ are also similar (not shown).
We will discuss this point in comparison with the results by the perturbation theory in Secs.~\ref{sec:perturb} and \ref{sec:field_dep}. 
Meanwhile, in the high-field FFM state, we find that ${\bf B}\parallel z$ gives rise to smaller tunnel spin currents than ${\bf B}\parallel c$ in both FM and AFM cases as shown in Fig.~\ref{fig:fig05}(c), 
reflecting the larger Zeeman gap observed in the spectra of $X_+(\omega)$ in Figs.~\ref{fig:fig05}(a) and \ref{fig:fig05}(b)~\footnote{
The results of FFM [Fig.~\ref{fig:fig05}(c)] show a behavior of $\tilde{J}_S \propto T$ at low temperatures, despite the Zeeman gap. 
This is a numerical artifact due to the finite $\Delta t$, 
which is not completely eliminated by the cubic spline interpolations. 
Nevertheless, we confirm that the effect on the results of KSL [Fig.~\ref{fig:fig04}(c)] is negligibly small due to the presence of low-energy excitations. 
}.
We will also discuss this issue in comparison with the results by the linear spin-wave theory in Secs.~\ref{sec:spinwave} and \ref{sec:field_dep}.

%%%%%%%%%%%%%%%%%%%%%%%%%%%%%%%%%%%%%%%%%%%%%%
%%%%%%%%%%%%%%%%%%%%%%%%%%%%%%%%%%%%%%%%%%%%%%
\subsection{Perturbation analysis} \label{sec:perturb}

\begin{figure}[htp]
\includegraphics[trim = 0 0 0 0, width=\columnwidth, clip]{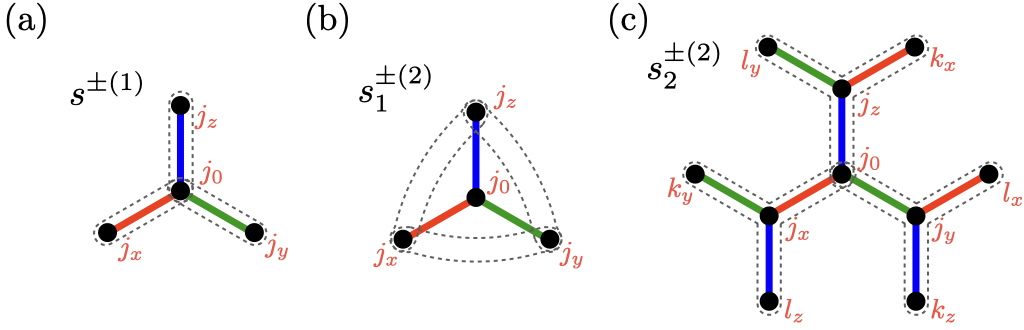}
\caption{\label{fig:fig06}
Pictorial descriptions of the effective operators
(a) $s^{\pm(1)}$,
(b) $s^{\pm(2)}_1$, and
(c) $s^{\pm(2)}_2$.
Each term is represented by the product of the Majorana operators at the 
two or four sites enclosed by the dashed lines.
}
\end{figure}

To examine the origin of the different sign of $\tilde{J}_S$ between the FM and AFM KSLs in Fig.~\ref{fig:fig04}(c), we calculate the tunnel spin current in terms of Majorana fermions by using the perturbation theory with respect to the magnetic field. 
Following the seminal paper by Kitaev~\cite{Kitaev2006},
we represent each spin $S=1/2$ operator by four Majorana fermion operators as
$S^\gamma_j = i b^\gamma_j c_j /2$, where $b^\gamma_j$ and $c_j$ are the Majorana fermion operators. 
By the transformation, 
the original spin degree of freedom is fractionalized into two types of quasiparticles: 
itinerant Majorana fermions and localized $Z_2$ fluxes.
The latter is defined by a product of the Majorana fermion operators $b_j^\gamma$, 
being local conserved quantities that take on values of $\pm 1$.
The sector with all $+1$ fluxes contains the ground state of the system and is called flux free sector.
In the spinless Majorana fermion representation,
the Kitaev Hamiltonian at zero field in the flux free sector is represented as
\begin{align}
\mathcal{H}_0 = - i \frac{K}{4} \sum_{\gamma} \sum_{\langle j,j' \rangle} c_j c_{j'}
\label{eq:H0}
\end{align}
with the spinless Majorana fermion operators $c_j$.
Based on the general framework of the perturbation theory~\cite{Takahashi1977},
we obtain the effective operator of $S^+_{j_0}$ in the flux free sector through a transformation 
by $\Gamma := PP_0 (P_0PP_0)^{-1/2}$, 
where $P_0$ and $P$ are projectors to the flux free sector of the unperturbed Hamiltonian [Eq.~\eqref{eq:H} for $B=0$] 
and to the states of the full Hamiltonian
that are adiabatically connected to those in the flux free sector at $B=0$, respectively.
By expanding up to the second order with respect to $B=|{\bf B}|$, 
the effective operator is obtained as
\begin{align}
\Gamma^\dag S^+_{j_0} \Gamma \sim s^{+(1)}+s^{+(2)}_1+s^{+(2)}_{2} =:s^{+}_{j_0},
\end{align}
where 
$s^{+(1)}$ and $s^{+(2)}_{1,2}$ are the first and second order  contributions with respect to 
$B$, respectively; $s^{+(1)}$ and $s^{+(2)}_{1}$ are quadratic with respect to the Majorana fermion operators, 
while $s^{+(2)}_{2}$ is quartic, as
\begin{align}
s^{+(1)}
\simeq& \frac{2}{|K|} \sum_\gamma B^\gamma A^\gamma 
\Big[ S_{j_0}^\gamma S_{j_\gamma}^\gamma + \big( S^\gamma_{j_0}\big)^2 \Big] \nonumber\\
=& \frac{1}{2|K|} \sum_\gamma B^\gamma A^\gamma \big( 1 -  i  c_{j_0} c_{j_\gamma} \big), \label{eq:sp1}\\
s^{+(2)}_1
\simeq& \frac{6}{|K|^2} \Big(
   A^x B^y B^z S^x_{j_0} S^y_{j_y} S^z_{j_z} \nonumber\\
 & +A^y B^z B^x S^y_{j_0} S^z_{j_z} S^x_{j_x} 
    +A^z B^x B^y S^z_{j_0} S^x_{j_x} S^y_{j_y} 
 \Big) \nonumber\\
=& -\frac{3 i }{4|K|^2} \Big(
   A^x B^y B^z  c_{j_y} c_{j_z} \nonumber\\
& +A^y B^z B^x c_{j_z} c_{j_x} 
 +A^z B^x B^y c_{j_x} c_{j_y} 
 \Big),\label{eq:sp21}\\
s^{+(2)}_2
\simeq& \frac{6}{|K|^2} \Big(
   A^x B^y B^z S^x_{j_0} S^y_{k_y} S^z_{l_z} \nonumber\\
 & +A^y B^z B^x S^y_{j_0} S^z_{k_z} S^x_{l_x} 
    +A^z B^x B^y S^z_{j_0} S^x_{k_x} S^y_{l_y} 
 \Big) \nonumber\\
=& - 
\frac{3 }{4|K|^2} 
\Big(
   A^x B^y B^z  c_{j_0} c_{j_x} c_{k_y} c_{l_z} \nonumber\\
& +A^y B^z B^x c_{j_0} c_{j_y} c_{k_z} c_{l_x}
 +A^z B^x B^y c_{j_0} c_{j_z} c_{k_x} c_{l_y}
 \Big). \label{eq:sp22}
\end{align}
Here, ${\bf B} = (B^x,B^y,B^z)$ and $S^+_{j_0}= A^x S^x_{j_0} + A^y S^y_{j_0} + A^z S^z_{j_0}$. 
See Fig.~\ref{fig:fig06} for the definition of lattice sites $j_\gamma$, $k_\gamma$, and $l_\gamma$. 
In the derivation,
the energy of all the intermediate states measured from the
ground state is set at $|K|$ for simplicity, as in Ref.~\cite{Kitaev2006}. 
The effective operator of $S^{-}_{j_0}$, $s^{-}_{j_0}=s^{-(1)}+s^{-(2)}_1+s^{-(2)}_{2}$, is given by replacing $A^\gamma$ by $(A^\gamma)^*$.
Then, the tunnel spin current is obtained by Eq.~\eqref{eq:chi-+}, where $S^{\pm}_{j_0}$ and $\mathcal{H}$ are replaced by $s^{\pm}_{j_0}$ and $\mathcal{H}_0$, respectively.
When we use the spectral representation, $X_+(\omega)$ in Eq.~\eqref{eq:X_pm} is written as
\begin{align}
 \left.X_+(\omega)\right|_{S^{\pm}_{j_0} \to s^{\pm}_{j_0}, \mathcal{H} \to \mathcal{H}_0} =
 \delta \sum_\ell 
\frac{
|\langle \ell | s^-_{j_0} |\emptyset \rangle |^2
-|\langle \ell | s^+_{j_0} |\emptyset \rangle |^2
}{(E_\ell -E_\emptyset - \omega)^2+\delta^2}
\label{eq:X_pm_spe}
\end{align}
for $\omega>0$,
where $|\emptyset \rangle$ is the ground state of the pure Kitaev model in the flux free sector~[Eq.~\eqref{eq:H0}],
$E_\emptyset$ is the ground state energy,
and
$|\ell \rangle$ and $E_\ell$ represent an excited state and its energy, respectively.
By straightforward calculations in terms of the Majorana operators, we find
that the $\mathcal{O}(B^2)$ contributions from $s^{+(1)}$ vanish in the numerator of Eq.~\eqref{eq:X_pm_spe}, 
since all the states appear in pairs satisfying
$|\langle \ell | s^{-(1)} |\emptyset \rangle | = |\langle \ell' | s^{+(1)} |\emptyset \rangle |$ with $E_\ell=E_{\ell'}$~(see Appendix~\ref{app:B}).
Note that there is no $\mathcal{O}(B)$ contribution either since $s^{\pm}$ does not contain $\mathcal{O}(B^0)$ term for the flux-free state.
Hence, the sum in Eq.~\eqref{eq:X_pm_spe} includes the $\mathcal{O}(B^3)$ and $\mathcal{O}(B^4)$ contributions. 
We therefore compute the lowest-order contribution of the tunnel spin current by
\begin{align}
 \tilde{J}_S \sim \frac{B^3}{|K|^3} \tilde{j}_S
\end{align}
with
\begin{align}
\tilde{j}_S 
&= 
\int_{-\infty}^{\infty}
d\omega \,
k
\left(
\frac{\omega}{T}
\right)\,
 x_{+}(\omega), 
\label{eq:js}\\
x_+(\omega) &= 
\lim_{B\to 0} 
\frac{|K|^3}{B^3}
 \left.X_+(\omega)\right|_{S^{\pm}_{j_0} \to s^{\pm}_{j_0}, \mathcal{H} \to \mathcal{H}_0}.
 \label{eq:x+}
\end{align}

We note that our perturbation analysis takes into account lower-order contributions than 
those in the third-order perturbation analysis 
in the seminal paper by A. Kitaev~\cite{Kitaev2006}.
Although the latter is known to open a gap in the spectra of itinerant Majorana fermions depending on the field direction, 
the effect of such gap opening is not included in our analysis; 
it appears in the higher-order $\mathcal{O}(B^5)$ contributions.

%%%%%%%%%%%%%%%%%%%%%%%%%%%%%%%%%%%%%%%%%%%%%%
\begin{figure}[htp]
\includegraphics[trim = 0 0 0 0, width=\columnwidth, clip]{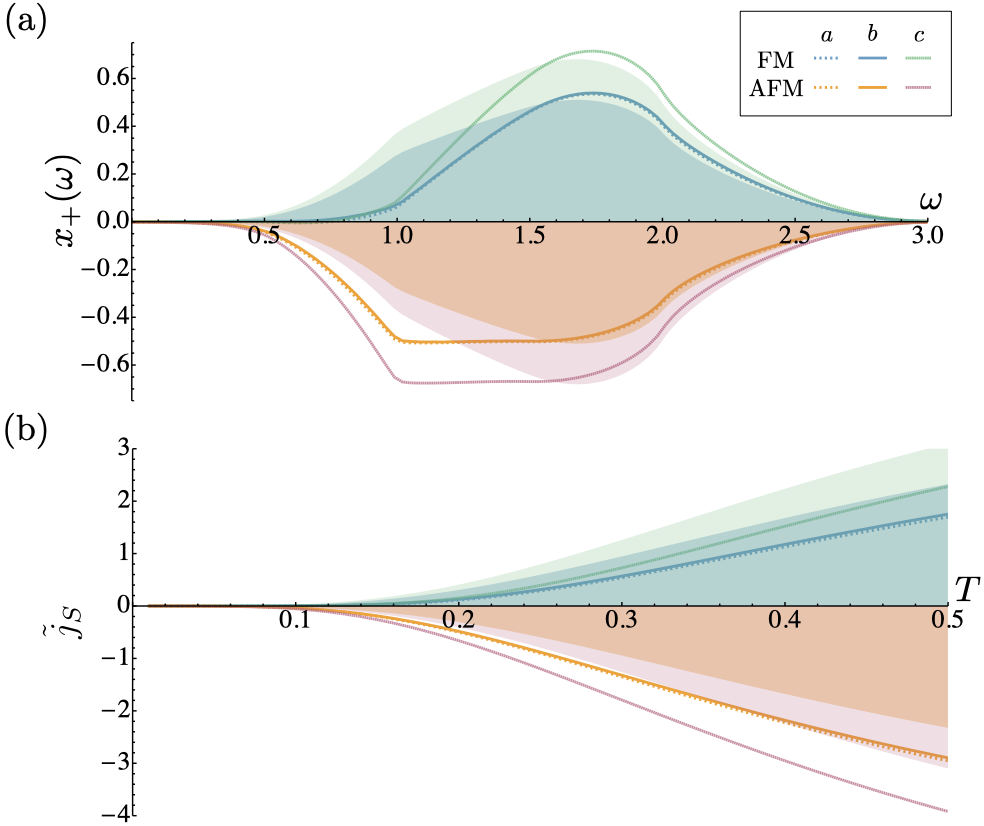}
\caption{\label{fig:fig07}
	Results of the perturbation analysis 
	for the FM and AFM Kitaev models in the weak magnetic field limit 
	with ${\bf B}\parallel a$, ${\bf B}\parallel b$, and ${\bf B}\parallel c$: 
	(a) $\omega$ dependences of $x_+(\omega)$ [Eq.~\eqref{eq:x+}] and 
	(b) temperature dependences of $\tilde{j}_S$ [Eq.~\eqref{eq:js}].
	The lines represent the full results including all the contributions from $s^{+(1)}$ and $s_{1,2}^{+(2)}$, 
	and the hatched regions represent the results ignoring $s^{\pm(2)}_{2}$. 
	In the latter hatched regions, ${\bf B}\parallel a$ and ${\bf B}\parallel b$ give the same results.
}
\end{figure}
%%%%%%%%%%%%%%%%%%%%%%%%%%%%%%%%%%%%%%%%%%%%%%

Figure~\ref{fig:fig07} shows the results of the perturbation analysis.
We find that $x_{+}(\omega)$ is 
mostly positive for the FM case (slightly negative at low $\omega$ for ${\bf B}\parallel a$ and ${\bf B}\parallel c$) and 
entirely negative for the AFM case, 
as shown in Fig.~\ref{fig:fig07}(a). 
This leads to positive and negative $\tilde{j}_S$ for the FM and AFM cases, respectively, as shown in Fig.~\ref{fig:fig07}(b). 
The result is in qualitative agreement with those of $\tilde{J}_S$ obtained 
by the TDVP simulations in Fig.~\ref{fig:fig04}(c).

The perturbation analysis provides a useful insight on the origin of the sign difference of the tunnel spin current between the FM and AFM cases.
When we neglect the four Majorana contributions of $s_2^{\pm(2)}$, one can show that 
$x_{+}(\omega)$ in the FM and AFM cases is identical in the absolute value, 
but the signs are opposite.
This is because the $\mathcal{O}(B^3)$ contributions in Eq.~\eqref{eq:X_pm_spe} 
from the cross terms between $ s^{\pm(1)}$ and $s^{\mp(2)}_1$ 
change their sign by $K\to -K$, i.e., 
$ \langle \emptyset| s^{\mp(2)}_1 | \ell  \rangle \langle \ell | s^{\pm(1)} |\emptyset \rangle \to -
 \langle \emptyset| s^{\mp(2)}_1 | \ell  \rangle \langle \ell | s^{\pm(1)} |\emptyset \rangle $, 
due to the sign change in the eigenstates of Majorana fermions (see Appendix~\ref{app:B}). 
Consequently, the interchange of FM and AFM $K$ causes a sign reversal of $\tilde{j}_S$ in Eq.~\eqref{eq:js}.
The contributions with neglecting $s_2^{\pm(2)}$ are plotted by the hatched regions in Fig.~\ref{fig:fig07}. 
The result indicates that the contributions 
from the quadratic terms of the Majorana operators,
$ s^{\pm(1)}$ and $s^{\pm(2)}_1$, are dominant in $\tilde{j}_S$, 
and moreover, are the main cause of the sign switching of the tunnel spin current between the FM and AFM cases. 
The asymmetry between the two cases is due to the quartic terms in $s_2^{\pm(2)}$,
leading to the faster growth of the tunnel spin current with temperature in AFM than FM, 
which is also consistent with the TDVP results in Fig.~\ref{fig:fig04}(c). 

Let us make a remark on the $B$ dependence of the tunnel spin current. 
The perturbation theory in the low-field limit predicts $\tilde{j}_S\propto B^3$, but $\tilde{J}_S$ obtained by TDVP appears not to follow this asymptotic form; 
we find that $\tilde{J}_S$ at $B=B_{\rm KSL}/2$ is about half of that at $B=B_{\rm KSL}$ (not shown), suggesting that $\tilde{J}_S$ is roughly proportional to $B$.
There are at least two possible reasons for this discrepancy. 
One is the contributions from excited states beyond the flux-free sector, which are neglected in the current perturbation analysis. 
The intermediate states [$|\ell \rangle$ in Eq.~\eqref{eq:X_pm_spe}] including flux excitations may give $\mathcal{O}(B)$ contributions, although they will be suppressed exponentially in the low-field limit because of the finite flux gap.
If this is the case, the perturbation theory predict $\tilde{j}_S\simeq p B^3 + q B e^{-\Delta_{\rm f}/T}
$, where $\Delta_{\rm f}$ is the flux gap $\simeq 0.06 |K|$~\cite{Kitaev2006}. 
The other possibility is the limited system sizes in the TDVP simulations. 
Larger size simulations allow us to distinguish the $B$ dependence by accessing lower-field 
regions with higher precision by taking smaller damping factor $\delta$ in Eq.~\eqref{eq:chi-+}. 
While such calculations require huge computational costs and are left for future study, 
this study comprehensively suggests that the coefficient $p$ and $q$ are both positive (negative) for the FM (AFM) KSL.

%%%%%%%%%%%%%%%%%%%%%%%%%%%%%%%%%%%%%%%%%%%%%%
\subsection{Linear spin-wave theory}\label{sec:spinwave}

\begin{figure}[!hpt]
\centering
\includegraphics[trim = 0 0 0 0, width=\columnwidth,clip]{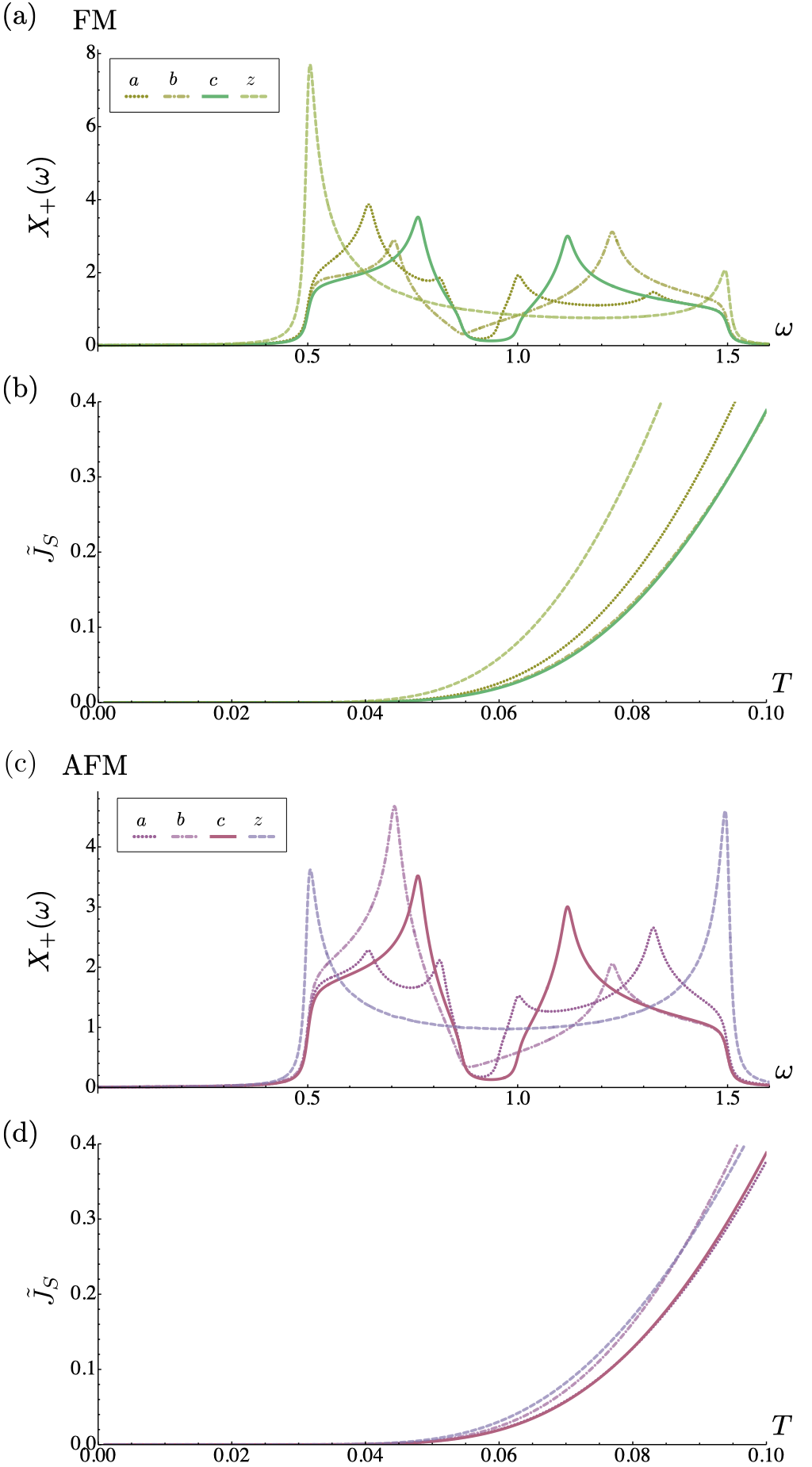}

\caption{\label{fig:fig08}
Results by the linear spin-wave theory for the (a,b) FM and (c,d) AFM Kitaev models at $B=B_{\rm FFM}$ 
with four field directions ${\bf B}\parallel a$, ${\bf B}\parallel b$, ${\bf B}\parallel c$, and ${\bf B}\parallel z$.
(a) and (c) show the $\omega$ dependences of $X_+(\omega)$ for the FM and AFM cases, respectively.
(b) and (d) show the corresponding temperature dependences of the tunnel spin current $\tilde{J}_S$.
}
\end{figure}

In this section, 
we present the results obtained by the linear spin-wave theory for the Kitaev model in the FFM state, for comparison with the results in Fig.~\ref{fig:fig05}. 
The procedure to calculate the tunnel spin current is described for the FM Heisenberg model in Appendix~\ref{app:A}. 
We take the similar procedure by performing the diagonalization of the bosonic Hamiltonian and the calculations of $X_+(\omega)$ numerically.

Figure~\ref{fig:fig08} shows the results for $B=B_{\rm FFM}$ (see Fig.~\ref{fig:fig03}) with four different field directions.
We find that $X_+(\omega)$ is always positive for both FM and AFM cases, 
as shown in Figs.~\ref{fig:fig08}(a) and \ref{fig:fig08}(c), respectively.
These behaviors are in qualitative agreement with the TDVP simulation results shown in Figs.~\ref{fig:fig05}(a) and \ref{fig:fig05}(b).
The positive $X_+(\omega)$ leads to the tunnel spin current with positive sign as shown in Figs.~\ref{fig:fig08}(b) and \ref{fig:fig08}(d). 
This is also in agreement with the TDVP simulation results in Fig.~\ref{fig:fig05}(c).
We note, however, that the gap in the spectrum of $X_+(\omega)$ is 
independent of the field directions in the linear spin-wave theory, 
while it is considerably different between ${\bf B}\parallel z$ and ${\bf B}\parallel c$ in the TDVP simulation.

In Figs.~\ref{fig:fig08}(a) and \ref{fig:fig08}(c), 
$X_+(\omega)$ shows a sharp peak at the lower edge of the spectrum for ${\bf B}\parallel z$ in both FM and AFM cases, 
while it shows a shoulder for the other field directions.
These qualitative differences can be understood as follows.
When ${\bf B}\parallel z$, the Holstein-Primakoff bosons cannot hop on the $z$ bonds and move only along one-dimensional zigzag chains composed of the $x$ and $y$ bonds, 
resulting in the logarithmic divergence in the spectrum. 
Meanwhile, for the other field directions,
two-dimensional motion is allowed, which results in the shoulder-like feature in the spectrum.
We note that these behaviors are not clearly seen in the TDVP simulation results~[Figs.~\ref{fig:fig05}(a) and \ref{fig:fig05}(b)], 
presumably due to finite-size effects and correlation between magnons.

%%%%%%%%%%%%%%%%%%%%%%%%%%%%%%%%%%%%%%%%%%%%%%

\section{Discussion}\label{sec:discussion}

\subsection{Sign of spin current in Kitaev spin liquids}

The real-time dynamics simulations based on TDVP reveal the sign switching of the 
tunnel spin current between the FM and AFM KSLs~(Sec. \ref{sec:realtime}). 
This behavior is successfully reproduced by the perturbation theory in the Majorana representation~(Sec.~\ref{sec:perturb}).
These findings indicate that the spin current in the KSL is predominantly carried by the itinerant Majorana fermions. 
More importantly, 
they suggest that the itinerant Majorana fermions contribute to the spin current 
with (up-)down-spin-like nature in the (AFM) FM KSL where the sign of the tunnel spin current is (negative) positive, 
even though they do not have spin angular momentum.

The negative SSE that we find for the AFM KSL appears to be common to that for a quantum spin liquid in a quasi-one-dimensional antiferromagnet~\cite{Hirobe2017}. 
This implies common physics in the 
spin current carried by fractional excitations in AFM quantum spin liquids in one and two dimensions, 
even though the fractional quasiparticles have different nature: 
The spin angular momentum carried by a spinon in the one-dimensional quantum spin liquid is half-integer quantized $\Delta S = 1/2$, while that by a Majorana fermion in the KSL is not. 
Our results, however, indicate that the FM quantum spin liquid, which is peculiar to the Kitaev model, can induce a tunnel spin current with positive sign, 
suggesting that the SSE is useful to identify the nature of fractional quasiparticles in quantum spin liquids. 
We emphasize that 
our study reveals, for the first time to our knowledge, that 
the SSE in quantum spin liquids can be either positive or negative
by analyzing
both FM and AFM Kitaev models in two dimensions.
Toward deeper understanding of the SSE in quantum spin liquids, 
further comparative studies are necessary for different types of quantum spin liquids in different dimensions. 
The simulation method for the real-time dynamics used in this study provides a powerful method to study generic models and awaits wider applications in the future.

Meanwhile, the SSE for the FM KSL has the same positive sign as that carried by conventional magnons in the FFM state. 
Thus, a way to distinguish between the two is needed. 
We will discuss this point in the next subsection.

\subsection{Magnetic field dependence of tunnel spin current} \label{sec:field_dep}

\begin{figure}[!hpt]
\centering

\includegraphics[trim = 0 0 0 0, width=\columnwidth,clip]{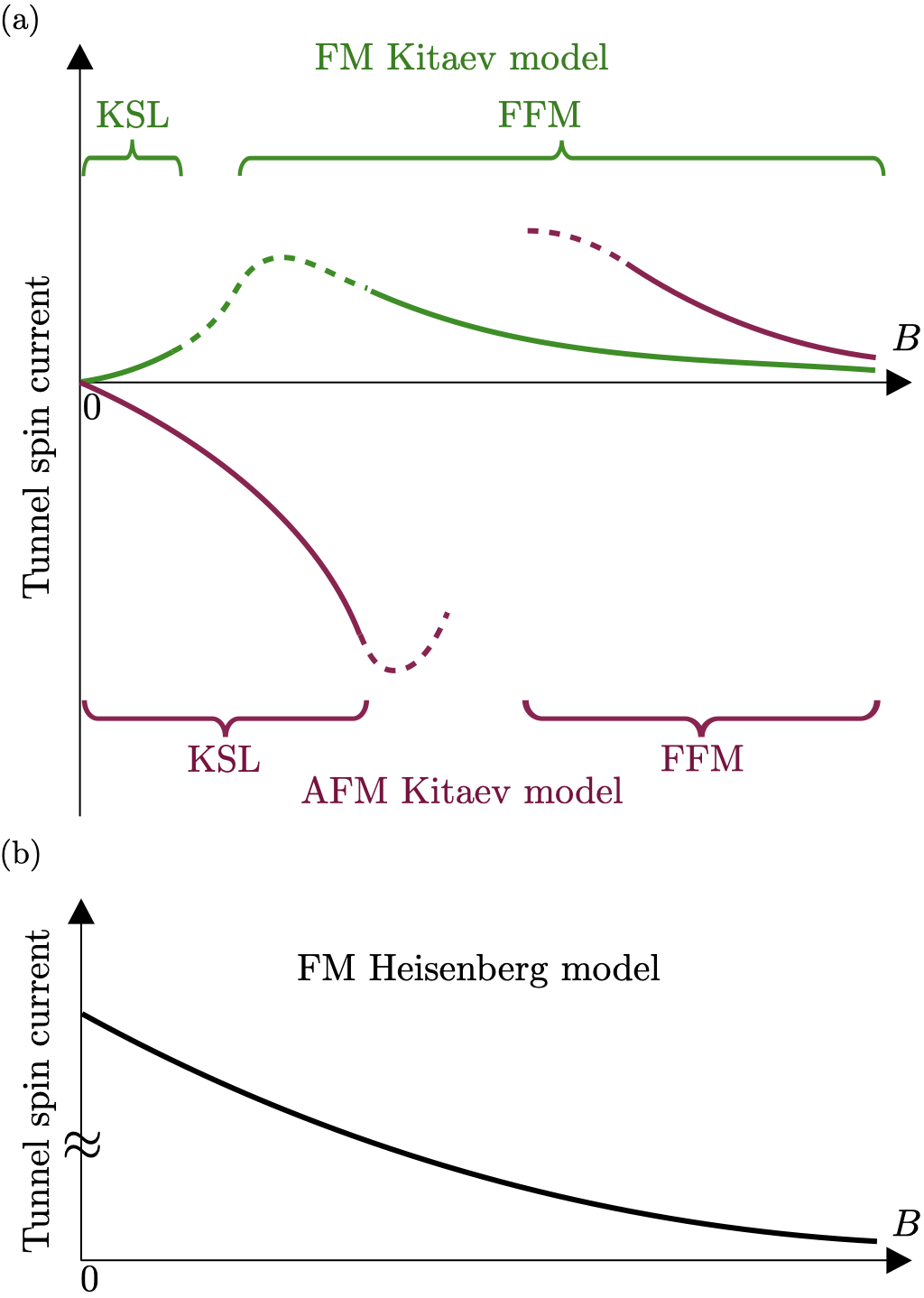}

\caption{\label{fig:fig09}
Schematics of the magnetic field dependence of the tunnel spin current
caused by the SSE at low temperature
for (a) the FM and AFM Kitaev models and (b) the FM Heisenberg model.
}
\end{figure}

Based on the results by the real-time dynamics simulations, the perturbation theory, and the linear spin-wave theory, 
we show a schematic of the magnetic field dependence of the tunnel spin current caused by the SSE for the Kitaev model in Fig.~\ref{fig:fig09}(a).
On one hand, 
in the low-field KSL region, the tunnel spin current increases in amplitude with increasing the field strength, with a different sign between the FM and AFM cases.
As discussed in the end of Sec.~\ref{sec:perturb}, a more detailed analysis is needed to fully elucidate the behavior in the low-field limit.
On the other hand, in the high-field FFM regime, the tunnel spin current is positive for both FM and AFM cases and decreases with increasing the field strength as shown in Fig.~\ref{fig:fig09}(a). 
This is similar to the FM Heisenberg model~[Fig.~\ref{fig:fig09}(b)].
Thus, these results imply that the tunnel spin current is maximized between the KSL and FFM states in the FM case, 
while it should change
its sign in the AFM cases, as schematically shown in Fig.~\ref{fig:fig09}(a). 
However, further studies are needed to clarify the field dependence in those transient regions, 
including the another intermediate SL region in the AFM case (see Fig.~\ref{fig:fig03}).

\begin{figure}[!hpt]
\centering

\includegraphics[trim = 0 0 0 0, width=\columnwidth,clip]{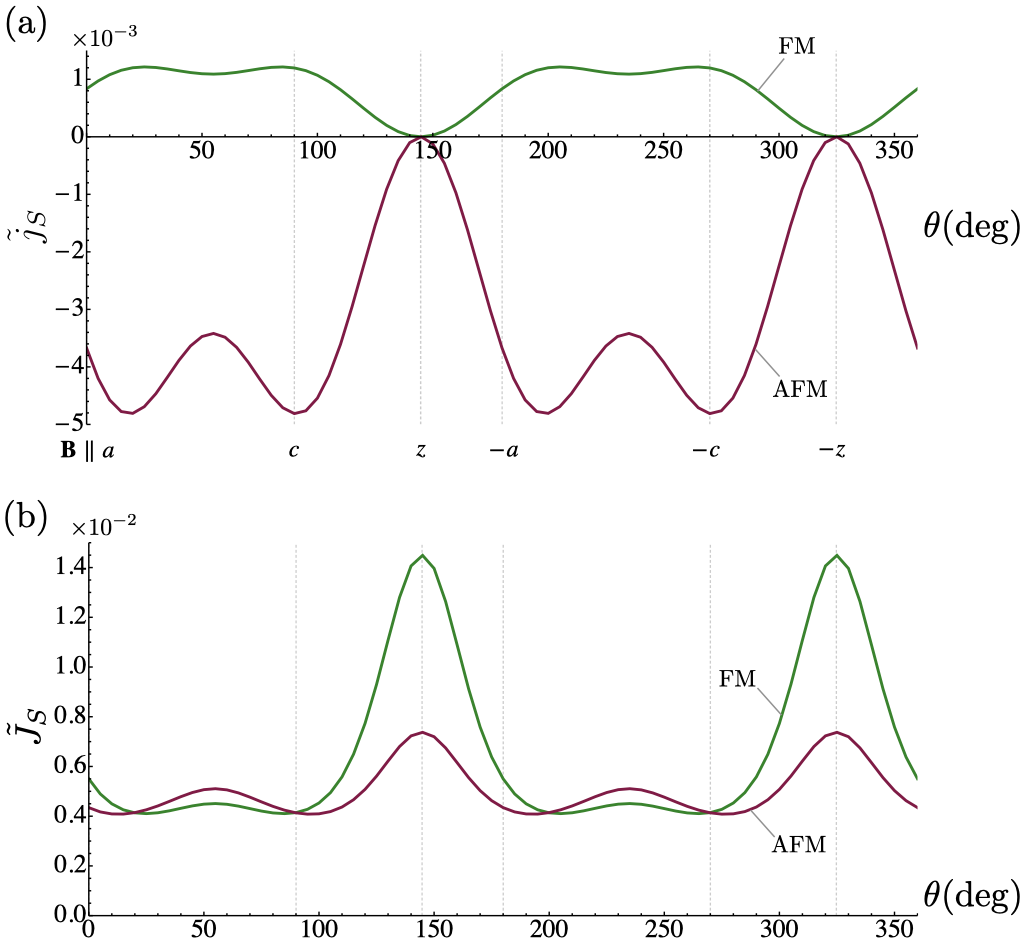}

\caption{\label{fig:fig10}
Field-angle dependences of the tunnel spin current when the magnetic field is rotated in the $a$-$c$ plane obtained by
(a) the perturbation analysis for low $B$
and (b) the linear spin-wave theory at $B=B_{\rm FFM}$. 
The angle $\theta$ is measured from the $a$ axis, and the data are calculated at $T=0.05$.
%In (a), the lines represent the full results including all the contributions from $s^{+(1)}$ and $s_{1,2}^{+(2)}$, and the hatched regions represent the results ignoring $s^{\pm(2)}_{2}$. 
}
\end{figure}

Since the positive sign of the SSE is common between the FM KSL and the FFM state, 
a way to distinguish between the two is needed. 
One is the field dependence discussed above: 
$\tilde{J}_S$ increases with $B$ in the low-field KSL, but it decreases in the high-field FFM state.
Another efficient way is the dependence on the magnetic field direction.
Figure~\ref{fig:fig10} shows the field angle dependence of the tunnel spin currents in the $ac$ plane~(see Fig.~\ref{fig:fig01}): 
(a) The results for the low-field KSL obtained by the perturbation analysis and (b) those for the high-field FFM state by the linear spin-wave theory. 
In the perturbation theory, the tunnel spin current $\tilde{j}_S$ vanishes for ${\bf B}\parallel z$. 
This is because $s^{\pm(1)}$, $s^{\pm(2)}_1$, and $s^{\pm(2)}_2$ in Eqs.~\eqref{eq:sp1}--\eqref{eq:sp22} are all zero for ${\bf B}\parallel z$, 
as $B^x=B^y=0$ and $A^z=0$. 
The same applies for ${\bf B}\parallel x$ and ${\bf B}\parallel y$ by symmetry.
Thus, the tunnel spin current for the low-field KSL is minimized for ${\bf B}\parallel z$, 
while maximized for ${\bf B}\parallel c$.
The other maxima with the same value as ${\bf B}\parallel c$ are found at $\theta \approx 20^\circ$ and $200^\circ$ 
due to the symmetry of $\pi$ rotation about the $[110]$ axis ($\theta \approx 54.7^\circ$).
Meanwhile, the linear spin-wave theory for the FFM state predicts an opposite tendency: 
The tunnel spin current is maximized for ${\bf B}\parallel z$, 
while minimized for ${\bf B}\parallel c$, as shown in Fig.~\ref{fig:fig10}(b).
This is due to the logarithmic divergence discussed in Sec.~\ref{sec:spinwave}. 
We note that the TDVP simulation data presented in Figs.~\ref{fig:fig04} and \ref{fig:fig05}, 
however, are not sufficient to discuss these issues presumably due to the limited system sizes. 
Further extensions are needed for more precise predictions for experiments.

\subsection{Experimental implication}
\label{sec:implication}

Our results of the SSE are expected to provide a strong insight into the identification of KSL in the candidate materials. 
As almost all the candidates show a magnetic order at low temperature at zero magnetic field due to parasitic non-Kitaev interactions such as the Heisenberg interaction, 
at least two scenarios have been proposed for the field dependence: 
One is that the system exhibits a direct transition from the low-field 
ordered phase to the FFM phase, 
and the other is that the system first exhibits a transition to an intermediate quantum spin liquid phase, 
and then at a higher field, it shows another transition or crossover to the FFM phase~
(see, e.g., Fig.~16 in Ref.~\cite{Trebst2022}). 
The inelastic neutron scattering (INS) measurements~\cite{Banerjee2017} as well as 
some of the thermal Hall conductivity measurements~\cite{Kasahara2018,Yamashita2020,Yokoi2021,Bruin2022} for the prime candidate $\alpha$-RuCl$_3$ support the latter scenario, 
even though the evidence remains inconclusive.
While our numerics has a limitation mainly because of the system size and the perturbation theory is limited to the weak-field region, 
we anticipate that our results contribute to understanding of the behavior under the magnetic field and the effect of the non-Kitaev interactions.

We emphasize that the field angle dependence of the SSE would be particularly important to distinguish the nature of quantum spin liquids. 
For instance, the RVB state, which is a prototypical example of quantum spin liquids, is expected not to exhibit nontrivial field-angle dependence unlike the KSL, 
since it is isotropic in spin space. 
Thus,  the SSE measurement in the intermediate-field region of the Kitaev candidate materials would serve as a good probe for not only  the existence 
but also the nature of the possible quantum spin liquid.

Besides the peculiar field-angle dependence, the sign change of the SSE signal would also be important in experiments for the Kitaev candidate materials. 
While the Kitaev interactions in the first-generation candidate materials,  
the Ir-based oxides $A_2$IrO$_3$ ($A$=Na, Li) and $\alpha$-RuCl$_3$,
are thought to be 
FM as theoretically predicted by Jackeli and Khaliullin~\cite{Jackeli2009}, 
there have been theoretical proposals for new generation,
such as cobaltites and $f$-electron compounds. 
In particular, it was predicted that the Kitaev interaction can be 
AFM in the $f$-electron systems~\cite{Jang2019,Jang2020}. 
Hence, our finding of the sign dependence of the SSE would be 
a useful experimental tool to identify the sign of the Kitaev interaction in the future materials exploration.

The SSE measurement represents a distinctive experimental technique, 
having several advantages over other methods used to study the Kitaev magnets
although it alone cannot fully unveil the nature of the KSL. 
The tunnel spin current is given by the frequency integral of the transverse component of the dynamical susceptibility, 
which is also measured by the INS using polarized neutrons~\cite{Knolle2014,Knolle2015,Banerjee2016,Yoshitake2016,Banerjee2017,Yoshitake2017,Yoshitake2017b,Do2017}. 
However, 
the SSE has advantages over the INS in several aspects: 
The INS requires a large-scale experimental facility and a huge amount of samples, and 
it is not suitable for some candidate materials when they include atoms that absorb neutrons (e.g., the Ir-based materials). 
In comparison, the SSE measurement can be done in much smaller-scale, table-top experiments, 
with a nano to micrometer scale sample, even in a monolayer. 
The electron spin resonance (ESR) is a method for detecting 
the spin dynamics at zero wavenumber in a certain frequency range,
and is complementary to the SSE measurement, 
in which the integral over the frequency and wavenumber is obtained~\cite{Ponomaryov2017}. 
The nuclear magnetic resonance (NMR) measures electron spin dynamics in the low-frequency limit 
and is qualitatively different from the information obtained by the SSE measurement~\cite{Jansa2018,Nagai2020}. 
The Raman scattering provides information over a wide frequency range,  
with the direction of polarization of the spin excitations depending on the polarization of light~\cite{Sandilands2015,Nasu2016,Wang2020}. 
This situation is similar in the SSE measurement, in which the polarization direction is identified by the direction of the magnetic field. 
Nevertheless, we emphasize that the SSE has an advantage as it can detect the sign of the Kitaev interaction, 
which is one of the main results of this study.
The thermal Hall effect and the SSE probe different nature of the Majorana excitations: The former observes their topological nature through the Berry curvature of the Majorana dispersions, while the latter observes their angular momenta.
We anticipate that a comprehensive approach, combining results from multiple experimental methods including the SSE  measurement, 
will shed further light on the intriguing phenomenon in the KSL.

\subsection{Experimental observability}
\label{sec:exp_obs}

In Sec.~\ref{sec:realtime}, we discussed that the tunnel spin current for the KSL is observable
in experiments from the comparison to the FM Heisenberg model on the two-dimensional honeycomb lattice. 
Here, we further reinforce the experimental observability
by additional comparison with a 3D model in a realistic experimental setup
as done in the previous studies~\cite{Adachi2011,Hirobe2017,Masuda2023}.
Specifically, we consider the $S=1/2$ Heisenberg model with the nearest-neighbor FM interaction $J$ under an external magnetic field $B$ on a simple cubic lattice, 
and calculate the tunnel spin current $\tilde{J}_S$ by the spin-wave theory in Appendix~\ref{app:A}. 
To evaluate the magnitude of $\tilde{J}_S$, 
we choose typical values of $B$ and $T$ that facilitate the observation in experiments.
Table~\ref{tab1} summarizes the calculated $\tilde{J}_S$ in the 3D FM state with 
$B/J=0.1$, $T/J=0.1$, and $0.5$, and those in the FM and AFM KSLs with $T=0.05$ [Fig.~\ref{fig:fig04}(c)]. 
Note that the critical temperature for the 3D FM state at $B=0$ is $T_C=0.839(1)$~\cite{Troyer2004,Wessel2010}.
These results suggest that the signal intensity in the KSLs is expected to be 
$\sim 10^{-2}$--$10^{0}$
of those in the typical 3D FM state.
Compared to the signal intensity of the spinon-mediated tunnel spin current observed in the previous experiment~\cite{Hirobe2017},
which is $10^{-4}$--$10^{-3}$
 times smaller than that of the 3D FM state,
the signal intensity in the KSLs is expected to be sufficiently large to be observed experimentally.
The signal in a typical 3D FM state is of the order of $1$ to $10~\mu\mathrm{V}$ in a typical setup~\cite{Kikkawa2023}.
Therefore, form these comparisons,
the signal intensity in the KSLs is roughly estimated to be in the range of $10~\mathrm{nV}$--$1~\mu\mathrm{V}$.

\begin{table}[!hbtp]
\caption{\label{tab1}
Comparison of the tunnel spin current $\tilde{J}_S$ between the KSL and typical 3D FM states.
The data for KSL are of $(X,L) = (6,10)$ and ${\bf B}\parallel c$ extracted from Fig.~\ref{fig:fig04}(c),
while those for the FM states are computed using the $S=1/2$ Heisenberg model on a simple cubic lattice
with $B/J=0.1$ based on the spin-wave theory (see Appendix~\ref{app:A}).
}
\begin{ruledtabular}
\begin{tabular}{llll}
System & State &  $T$  & $\tilde{J}_S$\\
\hline
\hline
3D Heisenberg & FM & $0.1 J$ & 0.178\\
3D Heisenberg & FM & $0.5 J$ & 4.73\\
\hline
2D FM Kitaev & KSL & $0.05 |K|$ & 0.111\\
2D AFM Kitaev & KSL & $0.05 |K|$ & -0.500\\
\end{tabular}
\end{ruledtabular}
\end{table}

%%%%%%%%%%%%%%%%%%%%%%%%%%%%%%%%%%%%

%%%%%%%%%%%%%%%%%%%%%%%%%%%%%%%%%%%%%%%%%%%%%%
\section{Summary}\label{sec:summary}
In summary, we have studied the spin Seebeck effect
in both the ferromagnetic and antiferromagnetic Kitaev spin liquids by using the real-time dynamics simulation, 
the perturbation analysis, and the linear spin-wave theory, based on the tunnel spin-current theory.
There are two main discoveries. 
The first one is that the spin Seebeck effect is induced in the Kitaev spin liquid by the low-energy fractional Majorana quasiparticles that have no spin angular momentum. 
This is a qualitatively different phenomenon from the spin Seebeck effect
in the one-dimensional quantum spin liquid induced by the fractional spinon excitations with nonzero spin angular momentum.
The second discovery is that the spin current mediated by elementary excitations in quantum spin liquids can be positive or negative; 
namely, we found the sign change of the spin current in the Kitaev spin liquid
between the ferromagnetic and antiferromagnetic Kitaev interactions.
Our finding suggests the Majorana fermions contribute to the spin current with up- and down-spin-like nature in the ferromagnetic and antiferromagnetic Kitaev spin liquids, respectively.
This finding refutes a possibility
that the spin current carried by the fractional excitations 
is always negative in the quantum spin liquids.

Fractional excitations in quantum spin liquids are hard to identify by conventional experimental probes, 
since they behave very differently from the original spin degree of freedom and there is no conjugate field to directly excite them.
Our finding suggests that the spin Seebeck effect would provide a strong complementary method
and would be a powerful tool for investigating the nature of fractional excitations, in addition to the existing methods, 
such as 
the thermal Hall effect~\cite{Kasahara2018,Yamashita2020,Yokoi2021,Bruin2022,Ye2018},
the inelastic neutron scattering~\cite{Knolle2014,Knolle2015,Banerjee2016,Yoshitake2016,Banerjee2017,Yoshitake2017,Yoshitake2017b,Do2017}, 
the Raman scattering~\cite{Sandilands2015,Nasu2016,Wang2020}, 
the electron spin resonance~\cite{Ponomaryov2017}, and 
the nuclear magnetic resonance~\cite{Jansa2018,Nagai2020}.

Our results show that the Majorana quasiparticles in the Kitaev spin liquid contribute to the spin current, 
suggesting the possibility of generating and controlling them by a spin injection.
It is highly desired to confirm this phenomenon experimentally by using the setup in Fig.~\ref{fig:fig01}(a) for both ferromagnetic and antiferromagetic Kitaev magnets; 
the former includes several iridium oxides and $\alpha$-RuCl$_3$~\cite{Takagi2019,Motome2020,Trebst2022}, 
and the latter was theoretically predicted, e.g., for polar spin-orbit Mott insulators~\cite{Sugita2020} and $f$-electron compounds~\cite{Jang2019,Jang2020}.

\begin{acknowledgments}
The authors would like to thank K. Fukui, H.-C. Jiang, K. Kobayashi, E. Saitoh, S. Trebst, and A. Tsukazaki for fruitful discussions.
This work was supported by Japan Society for the Promotion of Science (JSPS) KAKENHI Grant 
No.~JP19H05825, 
No.~JP20H00122, 
No.~JP20H01830,
No.~JP20H01849, 
No.~JP22H01179,
No.~JP22K03509,
No.~JP23H03818, No. JP23K22450, and No. JP24K00563,
JSPS Grant-in-Aid for Scientific Research on Innovative Areas
Grants 
No.~JP19H05825, 
No.~JP22H04480, 
No.~JP22H05131, and 
No.~JP23H04576,
JST COI-NEXT Program Grant Number JPMJPF2221,
and JST, CREST Grant No. JP-MJCR18T2  and No. JPMJCR24R5, Japan.
This work was also supported by the National Natural Science Foundation of China (Grant No.~12150610462).
Numerical calculations were performed using the facilities of the Supercomputer Center, The Institute for Solid State Physics, The University of Tokyo.
\end{acknowledgments}

%%%%%%%%%%%%%%%%%%%%%%
\renewcommand{\thesection}{\Alph{section}}
\renewcommand{\thesubsection}{\Alph{section}.\arabic{subsection}}
\renewcommand{\thesubsubsection}{\Alph{section}.\arabic{subsection}.\arabic{subsubsection}}
%\renewcommand{\thefigure}{S\arabic{figure}}
%\renewcommand{\theequation}{S\arabic{equation}}
%\renewcommand{\thetable}{S\Roman{table}}
%\baselineskip=6mm
%%%%%%%%%%%%%%%%%%%%%%
\appendix
%%%%%%%%%%%%%%%%%%%%%%%%%%%%%%%%%%%%%%%%%%%%%%
\section{Spin current in the ferromagnetic Heisenberg model}\label{app:A}

In this Appendix,
we show how to compute the tunnel spin current $\tilde{J}_S$ in Eq.~\eqref{eq:Js} for the $S=1/2$ FM Heisenberg model~\cite{Adachi2011,Hirobe2017,Masuda2023} shown in Fig.~\ref{fig:fig04}(c).
The Hamiltonian reads
\begin{align}
\mathcal{H}_{\rm H} = -J
\sum_{\langle j,j' \rangle} {\bf S}_j \cdot {\bf S}_{j'}
 - \sum_j  {\bf B}\cdot {\bf S}_j,
\label{eq:HH}
\end{align}
where the sum of $\langle j ,j' \rangle$ is taken for all the nearest-neighbor spins on the honeycomb lattice; 
we take $J = 1$.
Without loss of generality, we consider ${\bf B}\parallel z$, for which all the spins are fully polarized to the $z$ direction in the ground state.
The following calculations are described using the model on a honeycomb lattice, but extensions to models on other lattices are straightforward. We show the results for a simple cubic lattice in Sec.~\ref{sec:exp_obs}.

We calculate $\tilde{J}_S$ for the fully-polarized (FP) state by using the linear spin-wave theory.
By the Holstein-Primakoff transformation,
$S^+_j = b^{\;}_j$, 
$S^-_j = b^\dag_j$, and 
$S^z_j =\frac{1}{2} - n_j$ with $n_j = b^\dag_j b^{\;}_j$,
the Hamiltonian is written by the boson operators $b^{\;}_{j}$ and $b^\dag_{j}$ as
\begin{align}
\mathcal{H}_{\rm H} =&
-\frac{J}{2}
\sum_{\langle j,j \rangle} 
\left[
(b^\dag_j b_{j'}
+{\rm h.c.}
)
- n_j - n_{j'}
\right] \nonumber \\
&-J\sum_{\langle j,j' \rangle}  n_j n_{j'}
+B\sum_j n_j ,
\label{eq:HH2}
\end{align}
up to a constant, where $B=|{\bf B}|$.
Ignoring the interaction term $n_j n_{j'}$ and 
using the Fourier transformation,
we obtain
\begin{align}
\mathcal{H}^{\rm LSW}_{{\rm H}} =&
-\sum_{\bf q} 
\left(
g_{\bf q}
b^\dag_{{\bf q}\rm B} b_{{\bf q}\rm A}
+{\rm h.c.}
\right)\nonumber \\
&+\big(\frac{3}{2}J+B \big)
\sum_{{\bf q},\alpha}  b^\dag_{{\bf q}\alpha}b_{{\bf q}\alpha},
\end{align}
where $g_{\bf q} =(J/2) ( e^{i{\bf q}\cdot{\bf a}_1}+e^{i{\bf q}\cdot{\bf a}_2}+1)$ with two translation vectors ${\bf a}_1$ and ${\bf a}_2$, and the index $\alpha$ denotes the sublattice $\mathrm{A}$ and $\mathrm{B}$.
This Hamiltonian is easily diagonalized as
\begin{align}
\mathcal{H}^{\rm LSW}_{{\rm H}} =&
\sum_{{\bf q}}\sum_{\sigma=\pm}\varepsilon_{{\bf q}\sigma} \eta^\dag_{{\bf q}\sigma}  \eta^{\;}_{{\bf q}\sigma}, %\\
\end{align}
where
\begin{align}
\varepsilon_{{\bf q}\sigma} =&
\sigma |g_{\bf q}| +\frac{3}{2}J+B.
\end{align}
We note that the excitation spectra $\varepsilon_{{\bf q}\sigma}$ has a gap of $\Delta = B$ at ${\bf q}=0$.

For the FP state, 
the symmetrized dynamical spin susceptibility $X_+(\omega)$ [Eq.~\eqref{eq:X_pm}] is represented by using the spectral representation as
\begin{align}
X_{+}(\omega)=&\frac{ \delta}{2}
\sum_{\ell} 
\frac{ |\langle \ell |
 S^-_{j_0}
|{\rm FP} \rangle |^2
-
| \langle \ell |
S^+_{j_0} 
|{\rm FP} \rangle |^2
}{
(E_{\ell} - E_{\rm FP}-\omega)^2 + \delta^2
} \nonumber\\
=&\frac{ \delta}{2}
\sum_{\ell} 
\frac{ |\langle \ell |
 S^-_{j_0}
|{\rm FP} \rangle |^2
}{
(E_{\ell} - E_{\rm FP}-\omega)^2 + \delta^2
},
\end{align}
for $\omega>0$,
where $|{\rm FP} \rangle$ 
and $E_{\rm FP}$ are the FP ground state and its energy, 
and $|\ell \rangle$ and $E_{\ell}$ are an excited state and its energy; 
the sum of $\ell$ runs over all the excited states.
In the linear spin-wave theory, $X_+(\omega)$ is expressed in terms of boson operators as
\begin{align}
X_{+}(\omega)
&=
\frac{ \delta}{2}
\sum_{\ell} 
\frac{ |\langle \ell |
 b^\dag_{j_0}
|{\rm FP} \rangle |^2
}{
(E_{\ell} - E_{\rm FP}-\omega)^2 + \delta^2
} \nonumber\\
&=\frac{\delta}{2N_{\rm u}}
\sum_{{\bf q},\sigma} 
\frac{1}{
(\varepsilon_{{\bf q}\sigma}-\omega)^2 + \delta^2
} \nonumber \\
&\xrightarrow[\delta \to 0]{}
\ \frac{ \pi}{2N_{\rm u}}
\sum_{{\bf q},\sigma} 
\delta(\omega - \varepsilon_{{\bf q}\sigma}),
\end{align}
where $N_{\rm u}$ is the number of unit cells and we use the fact that $|{\rm FP} \rangle$ is a vacuum of bosons.
Finally, using this expression, the tunnel spin current is obtained as
\begin{align}
 \tilde{J}_{S} =& 2
 \int_{0}^\infty d\omega \,
X_{+} (\omega) \, 
k
\left(
\frac{\omega}{T}
\right)
=
\frac{\pi}{N_{\rm u}} \sum_{{\bf q},\sigma}
 k 
  \left(
  \frac{ \varepsilon_{{\bf q}\sigma} }{T} 
  \right).
\end{align}
In the calculation of the data in Figs.~\ref{fig:fig04}(c) and \ref{fig:fig05}(c), we take $N_{\rm u} = 288^2$.

%%%%%%%%%%%%%%%%%%%%%%%%%%%%%%%%%%%%%%%%%%%%%%
\section{Matrix elements of $s^{\pm(1)}$ and $s^{\pm(2)}_1$}\label{app:B}
In this Appendix,
we show two relations for the matrix elements of $s^{\pm(1)}$ and $s^{\pm(2)}_1$ in Eq.~\eqref{eq:X_pm_spe} used in the perturbation analysis.
Let us begin with diagonalization of the Majorana Hamiltonian for the flux free sector.
By using the Fourier transformation
\begin{align}
c_{{\bf r}\alpha} =& \sqrt{\frac{2}{N_{\rm u}}} {\sum_{\bf q}}' \left(
\gamma_{{\bf q}\alpha} e^{+i {\bf q}\cdot {\bf r}}
+
\gamma^\dag_{{\bf q}\alpha} e^{-i {\bf q}\cdot {\bf r}}\right),
\end{align}
where the sum runs over half of the first Brillouin zone with $q_x>0$ and $\alpha=\mathrm{A}$ or $\mathrm{B}$, 
Eq.~\eqref{eq:H0} is rewritten as
\begin{align}
\mathcal{H}_0 =& {\sum_{\bf q}}' 
{\boldsymbol \gamma}^\dag_{\bf q}
\mathsf{H}_{\bf q}
{\boldsymbol \gamma}^{\;}_{\bf q}
\end{align}
with
\begin{align}
\mathsf{H}_{\bf q}=&
\left(
\begin{matrix}
0 & if_{\bf q} \\
-if^{*}_{\bf q} & 0 
\end{matrix}
\right),
\end{align}
where
${\boldsymbol \gamma}^\dag_{\bf q}= ( \gamma^\dag_{{\bf q}w}~\gamma^\dag_{{\bf q}b} )$ 
and
$f_{\bf q} =(K/2) ( e^{i{\bf q}\cdot{\bf a}_1}+e^{i{\bf q}\cdot{\bf a}_2}+1)$.
The matrix $\mathsf{H}_{\bf q}$ is diagonalized by a unitary matrix,
\begin{align}
\mathsf{U}_{\bf q} = \frac{1}{\sqrt{2} |f_{\bf q}|}
\left(
\begin{matrix}
|f_{\bf q}| & if_{\bf q} \\
-if^{*}_{\bf q} & -|f_{\bf q}| 
\end{matrix}
\right)
\label{eq:Uq}
\end{align}
as
$\mathsf{U}_{\bf q}^\dag\mathsf{H}_{\bf q}\mathsf{U}_{\bf q}={\rm diag} ( |f_{\bf q}|, -|f_{\bf q}|)$.
Then, the Hamiltonian is diagonalized as
\begin{align}
\mathcal{H}_0 =& {\sum_{\bf q}}' 
\sum_{\sigma=\pm}
\sigma |f_{\bf q}|
\alpha^\dag_{{\bf q}\sigma}
\alpha^{\;}_{{\bf q}\sigma},
\end{align}
where ${\boldsymbol \alpha}^\dag_{\bf q} = (\alpha^\dag_{{\bf q}+}~\alpha^\dag_{{\bf q}-})$
and ${\boldsymbol \alpha}^\dag_{\bf q}  = {\boldsymbol \gamma}^\dag_{\bf q} \mathsf{U}_{\bf q}$.

With these notations, the ground state $|\emptyset \rangle $ is expressed as
\begin{align}
|\emptyset \rangle = {\prod_{\bf q}}' \alpha^\dag_{{\bf q}-} |{\rm vac}\rangle,
\end{align}
where $|{\rm vac}\rangle$ is the vacuum for $\alpha$ fermions.
Since the operators 
$s^{\pm(1)}$ and
$s^{\pm(2)}_1$ [Eqs.~\eqref{eq:sp1} and \eqref{eq:sp21}]
 have the quadratic forms of the Majorana fermion operators, 
 we consider two-particle, two-hole, and particle-hole 
states as the intermediate states $|\ell\rangle$:
$|{\bf q}+,{\bf q}'+\rangle = \alpha^\dag_{{\bf q}+} \alpha^\dag_{{\bf q}'+} |\emptyset \rangle$,
$|{\bf q}-,{\bf q}'-\rangle = \alpha_{{\bf q}-} \alpha_{{\bf q}'-} |\emptyset \rangle$,
and
$|{\bf q}+,{\bf q}'-\rangle = \alpha^\dag_{{\bf q}+} \alpha_{{\bf q}'-} |\emptyset \rangle$,
respectively.

First, we show that all the intermediate states appear in pairs of $|\ell\rangle$ and $|\ell'\rangle$, satisfying $|\langle \ell | s^{-(1)} |\emptyset \rangle | = |\langle \ell' | s^{+(1)} |\emptyset \rangle |$ with $E_\ell=E_{\ell'}$.
Let us consider the two-particle and two-hole states, for which
the matrix elements are given by
\begin{align}
&\langle {\bf q}-,{\bf q}'-| s^{+(1)} | \emptyset \rangle= 
-\frac{1}{2 N_{\rm u} |K| } \times
\nonumber\\
&
\left[\frac{f_{\bf q}}{ |f_{\bf q}|} 
 (B^x A^x e^{{ i}{\bf q}'\cdot {\bf a}_1} 
+ B^y A^y e^{{ i}{\bf q}'\cdot {\bf a}_2} 
+ B^z A^z) -({\bf q} \leftrightarrow {\bf q}')\right],
\label{eq:s1qmqm}
\end{align}
\begin{align}
&\langle {\bf q}'+,{\bf q}+| s^{-(1)} | \emptyset \rangle=
-  \frac{1}{2 N_{\rm u}|K|}\times \nonumber\\
&\biggl[
\frac{f_{{\bf q}'}}{|f_{{\bf q}'}|}
(B^x (A^x)^* e^{-i{\bf q}'\cdot {\bf a}_1}
+ B^y (A^y)^* e^{-i{\bf q}'\cdot {\bf a}_2}
+ B^z (A^z)^*) \nonumber \\
&\ \ -
({\bf q} \leftrightarrow {\bf q}')
\biggr],
\label{eq:s1qpqp}
\end{align}
respectively, where $({\bf q} \leftrightarrow {\bf q}')$ 
denotes the term with interchange of ${\bf q}$ and ${\bf q}'$.
From these two equations, we can show the relation
\begin{align}
\langle {\bf q}-,{\bf q}'-| s^{+(1)} | \emptyset \rangle^*= 
\langle {\bf q}'+,{\bf q}+| s^{-(1)} | \emptyset \rangle \frac{f^*_{\bf q} f^*_{{\bf q}'}}{|f^*_{\bf q} ||f^*_{{\bf q}'}|}.
\end{align}
Thus, we obtain
\begin{align}
|\langle {\bf q}-,{\bf q}'-| s^{+(1)} | \emptyset \rangle |= 
|\langle {\bf q}'+,{\bf q}+| s^{-(1)} | \emptyset \rangle |.
\label{eq:--++}
\end{align}
Note that the energy of $|\ell \rangle = |{\bf q}-,{\bf q}' -\rangle$ and $|\ell' \rangle = |{\bf q}+,{\bf q}'+ \rangle$ is the same, 
since $E_\ell = E_{\ell'} = |f_{\bf q}|+|f_{{\bf q}'}|$.
For the particle-hole states, the matrix element is given by
\begin{align}
&\langle {\bf q}+,{\bf q}'- | s^{+(1)}  |\emptyset \rangle=
  \frac{i}{2N_{\rm u}|K|} \times \nonumber\\
&\frac{1}{|f_{\bf q}||f_{{\bf q}'}|}
\biggl[
 (B^x A^x e^{+i{\bf q}'\cdot {\bf a}_1} 
+ B^y A^y e^{+i{\bf q}'\cdot {\bf a}_2} 
+ B^z A^z) 
|f_{\bf q}||f_{{\bf q}'}| \nonumber\\
&-
  (B^x A^x e^{-i{\bf q}\cdot {\bf a}_1}
+ B^y A^y e^{-i{\bf q}\cdot {\bf a}_2}
+ B^z A^z) 
 f_{{\bf q}} f_{{\bf q}'}
\biggr].\label{eq:s1qpqm}
\end{align}
Similarly to the two-particle and two-hole states,
we can show 
\begin{align}
\langle {\bf q}+,{\bf q}'-| s^{+(1)} | \emptyset \rangle^*= 
\langle {\bf q}'+,{\bf q}-| s^{-(1)} | \emptyset \rangle \frac{f^*_{\bf q} f^*_{{\bf q}'}}{|f^*_{\bf q} ||f^*_{{\bf q}'}|},
\end{align}
and thus, we obtain
\begin{align}
|\langle {\bf q}+,{\bf q}'- | s^{+(1)}  | \emptyset \rangle|
=|\langle {\bf q}'+,{\bf q}- | s^{-(1)}  |  \emptyset \rangle|.
\label{eq:+-+-}
\end{align}
In this case also, the excited states $|{\bf q}+,{\bf q}' -\rangle$ and $|{\bf q}-,{\bf q}'+ \rangle$ have the same energy. 
Equations~\eqref{eq:--++} and \eqref{eq:+-+-} prove that all the intermediate states are paired and satisfy $|\langle \ell | s^{-(1)} |\emptyset \rangle | = |\langle \ell' | s^{+(1)} |\emptyset \rangle |$ with $E_\ell=E_{\ell'}$.

Next, we show that $\langle \emptyset| s^{\mp(2)}_1 | \ell  \rangle \langle \ell | s^{\pm(1)} |\emptyset \rangle$ changes the sign by the sign reversal of $K$.
The matrix elements of $s^{+(2)}_1$ are expressed as
\begin{align}
&\langle {\bf q}-,{\bf q}'- | s^{+(2)}_{1}| \emptyset \rangle=
  \frac{3 i}{4N_{\rm u}|K|^2}\times \nonumber\\
& \biggl[
\left(
A^z B^xB^y 
e^{ i ({\bf q}\cdot {\bf a}_2+{\bf q}'\cdot {\bf a}_1)} 
+
A^x B^yB^z 
e^{ i {\bf q}'\cdot {\bf a}_2} 
+
A^y B^zB^x 
e^{ i {\bf q}\cdot {\bf a}_1} 
\right) \nonumber\\
&\ \ -({\bf q} \leftrightarrow {\bf q}')
\biggr],\\
%%%%%%%%%%%%
&\langle {\bf q}+,{\bf q}'+| s^{+(2)}_{1} |\emptyset \rangle=
 - \frac{3 i }{4N_{\rm u}|K|^2}   
 \times \nonumber\\
&\frac{f_{\bf q}  f_{{\bf q}'}}{|f_{\bf q}||f_{{\bf q}'}|} \biggl[
\Bigl(
A^z B^xB^y 
 e^{- i ({\bf q}\cdot {\bf a}_2 + {\bf q}'\cdot {\bf a}_1)}
+
A^x B^yB^z  e^{- i {\bf q}'\cdot {\bf a}_2}\nonumber\\
&\quad\quad+
A^y B^zB^x  e^{- i {\bf q}\cdot {\bf a}_1}
\Bigr)
-({\bf q} \leftrightarrow {\bf q}')
\biggr],\\
%%%%%%%%%%%%%
&\langle {\bf q}+,{\bf q}'- | s^{+(2)}_1 |\emptyset \rangle=
\frac{3}{4N_{\rm u}|K|^2}\times\nonumber\\
&
\frac{
f_{\bf q}
}{|f_{\bf q}|}
\biggl[
\Bigl(
A^z B^xB^y 
  e^{- i ({\bf q}\cdot {\bf a}_2 - {\bf q}'\cdot {\bf a}_1)}
+A^x B^yB^z 
 e^{ i {\bf q}'\cdot {\bf a}_2}\nonumber\\
&\quad\quad+ A^y B^zB^x 
  e^{- i {\bf q}\cdot {\bf a}_1}
\Bigr) 
- ({\bf q}' \leftrightarrow -{\bf q})
\biggr].
\end{align}
The matrix elements for $s^{-(2)}_1$ are obtained by replacing $A^\gamma$ by $(A^\gamma)^*$.
If the sign of $K$ is reversed, $f_{\bf q}$ becomes $-f_{\bf q}$,
and then,
$\langle {\bf q}\pm,{\bf q}'\pm| s^{\pm(1)} | \emptyset \rangle$
and
$\langle {\bf q}+,{\bf q}'-| s_1^{\pm(2)} | \emptyset \rangle$
change their signs,
while $\langle {\bf q}+,{\bf q}'-| s^{\pm(1)} | \emptyset \rangle$
and
$\langle {\bf q}\pm,{\bf q}'\pm| s_1^{\pm(2)} | \emptyset \rangle$ are intact.
 Therefore, 
 the sign reversal of $K$ leads to
 $\langle \emptyset| s^{\mp(2)}_1 | \ell  \rangle \langle \ell | s^{\pm(1)} |\emptyset \rangle
 \to
 -\langle \emptyset| s^{\mp(2)}_1 | \ell  \rangle \langle \ell | s^{\pm(1)} |\emptyset \rangle$.

%%%%%%%%%%%%%%%%%%%%%%

\bibliography{draft} 

\end{document}